\documentclass[showpacs,pre,aps,twocolumn,superscriptaddress]{revtex4-1}

\pdfoutput=1
\usepackage{amsmath,amsfonts,amssymb,bm,graphicx,color}

\usepackage{hyperref}
\usepackage{color}

\newcommand{\Q}{\mathcal{Q}}
\newcommand{\V}{\mathcal{V}}

\newcommand{\J}{\mathcal{J}}
\newcommand{\I}{\mathcal{I}}
\newcommand{\hQ}{\hat{\mathcal{Q}}}

\newcommand{\Tr}{{\rm Tr} \,}
\newcommand{\U}{\mathcal{U}}
\newcommand{\F}{\mathcal{F}}
\renewcommand{\S}{\mathcal{S}}
\renewcommand{\L}{\mathcal{L}}
\newcommand{\R}{\mathcal{R}}

\newcommand{\K}{{\mathbf K}}
\newcommand{\KK}{{\mathbb K}}
\newcommand{\JJ}{{\mathbf V}}
\newcommand{\TT}{{\mathbb T}}
\newcommand{\FF}{{\mathbb F}}
\newcommand{\II}{{\mathbb I}}

\newcommand{\T}{\mathcal{T}}
\newcommand{\hT}{\hat{\mathcal{T}}}

\newcommand{\KKK}{{\mathcal K}}

\newcommand{\er}[1]{Eq.~\eqref{#1}}
\newcommand{\ers}[2]{Eqs.~(\ref{#1}-\ref{#2})}
\newcommand{\era}[2]{Eqs.~(\ref{#1}) and (\ref{#2})}

\newcommand{\Ers}[2]{Equations~(\ref{#1}-\ref{#2})}

\begin{document}

\title{Catching and reversing quantum jumps and thermodynamics of quantum trajectories}

\author{Juan P. Garrahan}
\affiliation{School of Physics and Astronomy, University of
Nottingham, Nottingham, NG7 2RD, UK}
\affiliation{Centre for the Mathematics and Theoretical Physics of Quantum Non-equilibrium Systems,
University of Nottingham, Nottingham NG7 2RD, UK}

\author{M\u{a}d\u{a}lin Gu\c{t}\u{a}}
\affiliation{School of Mathematical Sciences, University of
Nottingham, Nottingham, NG7 2RD, UK}
\affiliation{Centre for the Mathematics and Theoretical Physics of Quantum Non-equilibrium Systems,
University of Nottingham, Nottingham NG7 2RD, UK}

\date{\today}

\begin{abstract}
A recent experiment by Minev et.\ al [arXiv:1803.00545] demonstrated that in a dissipative (artificial) 3-level atom with strongly intermittent dynamics it is possible to ``catch and reverse'' a quantum jump ``mid-flight'': by the conditional application of a unitary perturbation after a fixed time with no jumps,  
the system was prevented from getting shelved in the dark state, thus removing the intermittency from the dynamics. Here we offer an interpretation of this phenomenon in terms of the dynamical large deviation formalism for open quantum dynamics. In this approach, intermittency 
is seen as the {\em first-order coexistence} of active and inactive {\em dynamical phases}. Dark periods are thus like {\em space-time bubbles} of the inactive phase in the active one. Here we consider a controlled dynamics via the (single - as in the experiment - or multiple) application of a unitary control pulse during no-jump periods. By considering the large deviation statistics of the emissions, we show that appropriate choice of the control allows to stabilise a desired dynamical phase and remove the intermittency. In the thermodynamic analogy, the effect of the control is to prick bubbles thus preventing the fluctuations that manifest phase coexistence. We discuss similar controlled dynamics in broader settings. 
\end{abstract}
\maketitle

\section{Introduction}

Open quantum systems evolve stochastically due to the action of the environment with which they interact. 
This stochastic evolution is the result of conditioning on events observed in the environment, such as the detection of photons emitted by a driven atom. 
A time record of such emissions is a {\em quantum measurement trajectory} corresponding to the time sequence of {\em quantum jumps} in the state of the system. That is, to the observable (and classical) quantum measurement trajectory corresponds an unobservable {\em quantum trajectory} (also called a quantum filter) of the state of the system. In the simplest case of weak coupling to a large bath, the quantum trajectories of such dissipative dynamics are those of a quantum Markov process. 
For reviews see \cite{Plenio1998,Belavkin1999,Breuer2002,Gardiner2004}.

A recent experiment by Minev et al.\ \cite{Minev2018} has provided a beautiful practical demonstration of quantum trajectories. It studied a superconducting artificial atom designed in a three level $V$ geometry, like the one sketched in Fig.\ 1(a). Conditions were such that emission dynamics was intermittent due to the system occasionally getting ``shelved'' in the non-emitting state, cf.\ Fig.\ 1(b). A feedback and control mechanism was then devised to prevent excursions into the dark subspace: a longer-than-typical period  with no emissions - after the last observed one - was taken as an ``advanced warning'' that the system was in the process of moving away from the bright subspace and into the dark one. Conditioned on this observation, a unitary perturbation was then applied designed to rotate the state away from the dark level. The effect was to remove the intermittency in the dynamics by preventing the system from leaving the bright subspace, a procedure called in Ref.~\cite{Minev2018} ``catching and reversing'' a quantum jump ``mid-flight''. The success of the controlled dynamics required the ability to detect emissions with very high efficiency in order to get an accurate estimation of the state of the system, something that was possible in the solid-state setting of Ref.\ \cite{Minev2018} in contrast to what is achievable in actual atomic systems. The experiment of Ref.~\cite{Minev2018} - together with other recent ones such as  \cite{Murch2013,Weber2014,Tan2015,Foroozani2016} illustrating the experimental accessibility to quantum trajectory information - is a remarkable demonstration of the applicability of quantum trajectory concepts \cite{Belavkin1990,Dalibard1992,Gardiner1992,Carmichael1993,Mabuchi1996,Wiseman2002,Handel2005,Gambetta2008,Wiseman2009,Gammelmark2013,Guevara2015} in open quantum systems.

In this note we consider the problem above from the point of view of {\em thermodynamics of  quantum trajectories} \cite{Garrahan2010}. This is a statistical mechanics approach to the dynamics of open quantum systems that aims to treat ensembles of trajectories just like standard equilibrium statistical mechanics treats ensembles of configurations. It thus generalises concepts such as order parameters, free-energies and thermodynamic phases to the ensemble of trajectories generated by a dynamics. This method can be formalised by applying to dynamics the techniques of large deviations (LDs) \cite{Eckmann1985,Touchette2009,Esposito2009,Gaspard2005} (i.e., the same mathematical tools used to define the standard equilibrium ensemble method of statistical mechanics). This approach was originally devised for classical systems and has been successfully employed in the study of a variety of dynamical problems, uncovering in many cases the existence of rich {\em dynamical phase behaviour}, as for example in glasses \cite{Merolle2005,Garrahan2007,Hedges2009,Speck2012b},  exclusion processes and driven systems \cite{Appert2008,Hurtado2011,Espigares2013,Jack2015,Karevski2017,Baek2017}, signalling networks \cite{Vaikuntanathan2014,Weber2015}, and protein folding \cite{Weber2013,Mey2014}. For a basic review of the classical dynamical LD approach and its extension to Markovian open quantum systems see \cite{Garrahan2018}.

From the thermodynamics of trajectories perspective, intermittency in the emission dynamics of an open quantum system is related to {\em dynamical phase coexistence} \cite{Garrahan2010,Ates2012,Lesanovsky2013}. Intermittent dynamics is a consequence of the existence of two distinct dynamical phases, an {\em active} phase where emissions are plentiful, and an {\em inactive} phase with low or no emissions. This dynamical phase structure is what is revealed by the LD approach. At conditions where intermittency occurs, {\em typical} trajectories of the dynamics are those that display coexistence and alternate between these two phases. By quantifying the statistical properties of all trajectories,  one finds that away from typical behaviour - i.e., when {\em rare} fluctuations give rise to atypical dynamics - rare trajectories that emit much more than average have different characteristics than rare trajectories that emit much less than average, each belonging to a distinct dynamical phase. 

The change in the nature of dynamical fluctuations from rare and active, through typical and intermittent, to rare but inactive, corresponds to the physics of a first-order phase transition (as say, liquid to vapour in a standard equilibrium static setting) but occurring in trajectory space: it is first-order as the two phases have distinct values of the order parameter that distinguishes them (for example their characteristic emission rate - cf.\ the difference in density between the liquid and vapour phases in the static analogy) and the interface between bright and light periods when intermittent - i.e., at coexistence - is sharp (cf.\ the sharp interface of vapour bubbles in a liquid). For a system with a finite state space, such as the three level one of Fig.\ 1(a), the transition cannot be sharp and is rounded off (a singular transition requires a large system), and the dynamical behaviour is thus one of a first-order {\em crossover}.

Here we use the above methods and ideas to study and interpret the {\em quantum jump reversal} control dynamics as the one of the experiment of Ref.\ \cite{Minev2018}. The reason why such control dynamics is effective is precisely the phase coexistence character of the dynamics. The fact that the interface between regimes belonging to the two phases is sharp allows an accurate identification of when the control perturbation needs to be  applied. Below we study this control dynamics using the tools of dynamical LD. We find that
that appropriate choices of the control operation and the time at which it is applied  allows to stabilise a desired dynamical phase (either active or inactive) and remove the intermittency. In a static thermodynamic analogy, the effect of the control is akin to scanning a system close to phase coexistence and when seeing an interface pricking it so as to prevent the formation of bubbles of the other phase.  

The paper is organised as follows. In Sec.\ II we introduce the basic three level model and we review the LD properties of its dynamics. In Sec.\ III we study a feedback-control dynamics similar to the one used in Ref.\ \cite{Minev2018} within the LD approach, showing how appropriate control pulses allow to select a dynamical phase and remove intermittency. In Sec.\ IV we generalise the protocol to allow for repeated unitary pulses within a no-jump period, which gives even more precise control on the observed dynamics, and implement this feedback scheme in an alternative Markovian fashion by means of a ancillary classical controller. In Sec.\ V we discuss the broader context of our results and give our conclusions.

\begin{figure*}[t!]
\begin{center}
\includegraphics[width=\textwidth]{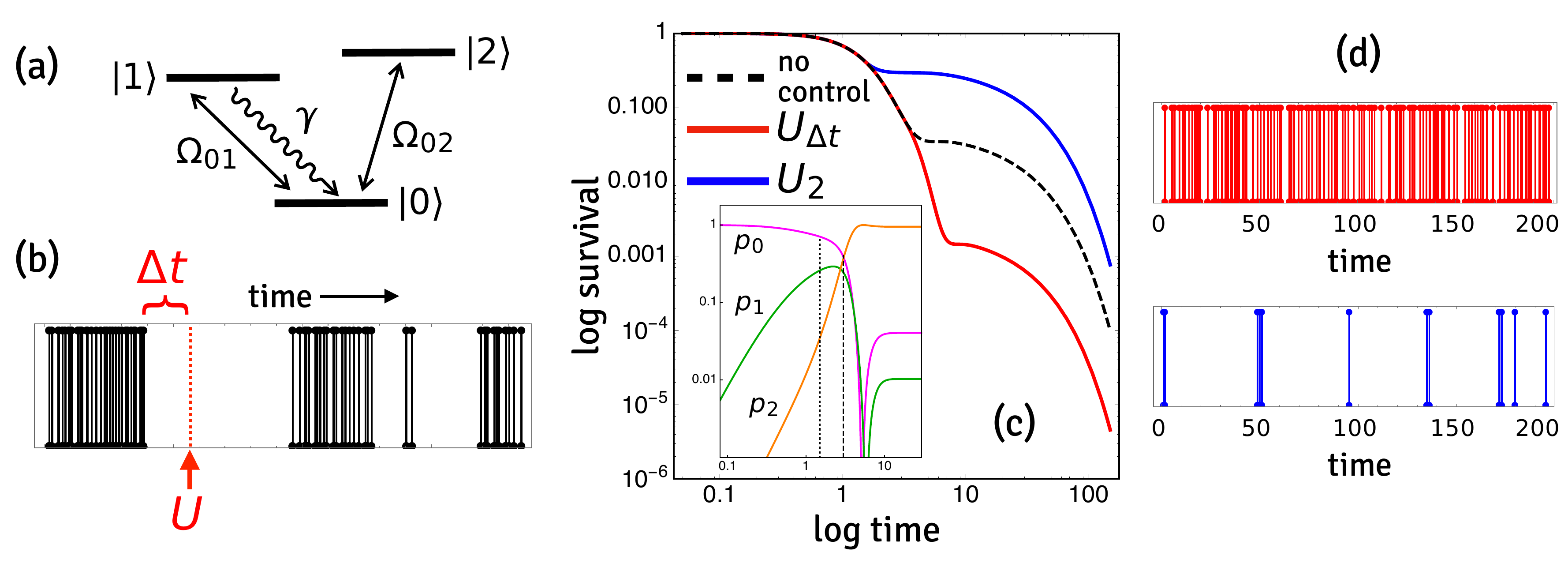}
\caption{
(a) Dissipative three level model. Levels $0$ and $1$ span the bright subspace, with level $2$ the shelving state. (In Ref.~\cite{Minev2018} states $0,1,2$ are called $G,B,D$, respectively.) 
(b) Typical emission trajectory and sketch of the control scheme. (c) Survival probabilities $S(t)$ for the original dynamics (black), for the controlled dynamics with $U_{\Delta t}$ and $\Delta t = 3$ (red), and for the controlled dynamics with $U_{2}$ and $\Delta t = 3/2$ (blue); time in units of $\Omega_{01}^{-1}$. Inset: probabilities of occupation $p_{0,1,2}$ of the states $|0,1,2\rangle$ as a function of time after the last renewal to $|0\rangle$. The dotted and dashed lines indicate the values of $\Delta t$ for the results in Figure 2. (d) Representative emission trajectories for the two control dynamics of Sec.\ IV, that with $U_{\Delta t}$ and $\Delta t = 3$ (red) at the top, and that with $U_2$ and $\Delta t = 1.5$ (blue) at the bottom. Parameters: $\Omega_{01}=1,\Omega_{02}=1/10,\gamma=4$. 
}
\end{center}
\end{figure*}

\section{Model and dynamical large deviations method}

\subsection{Three level system}

As in Ref.\ \cite{Minev2018}, we consider a system with three levels, $\{ | 0 \rangle, | 1 \rangle, | 2 \rangle \}$, in a $V$ setting, see Fig.~1. The dynamics we consider is Markovian, with the average state $\rho_t$ is given by 
\footnote{We indicate operators in the system with capital letters such as $J$, and super-operators on the system by calligraphic font symbols such as $\V$. We use bold symbols such as $\JJ$ and blackboard ones such as $\TT$ to indicate operators and super-operators, respectively, in the system-controller combined system of Sec.\ IV.B.}
\begin{equation}
\rho_t = \T_{0,t} \rho_0
\label{V}
\end{equation}
Here $\T_{t_1,t_2}$ corresponds to the evolution super-operator that evolves the average state between times $t_1$ and $t_2$. For the case where the dynamics is not controlled (see below for the controlled case) average evolution is generated by the Master super-operator $\L$, that is,
\begin{equation}
\T_{t_1,t_2} = e^{(t_2-t_1) \L} ,
\label{VL}
\end{equation}
with $\L$ of the Lindblad form \cite{Lindblad1976,Gorini1976}
\begin{equation}
\L (\cdot) = -i [ H, (\cdot) ] + J (\cdot) J^\dagger - \frac{1}{2} \{ J^\dagger J, (\cdot) \}
\label{L}
\end{equation}
with Hamiltonian and jump operator given by 
\begin{equation}
H = \Omega_{01} \left( | 0 \rangle \langle 1 | + | 1 \rangle \langle 0 | \right)
+ \Omega_{02} \left( | 0 \rangle \langle 2 | + | 2 \rangle \langle 0 | \right)
\label{H}
\end{equation}
and 
\begin{equation}
J = \sqrt{\gamma} ~ | 0 \rangle \langle 1 |
\label{J}
\end{equation}
This means that the evolution equation \er{V} can be written in the form of a Master equation 
\begin{equation}
\partial_t \rho_t = \L \rho_t .
\label{ME}
\end{equation}

The above dynamics is related to a quantum Markov process \cite{Breuer2002,Gardiner2004}, corresponding to an unravelling of the average dynamics in terms of stochastic quantum trajectories \cite{Belavkin1990,Dalibard1992,Gardiner1992,Carmichael1993}. Under conditions such as $\gamma,\Omega_{01} \gg \Omega_{02}$ this stochastic dynamics is {\em intermittent}, see. Fig.~1(b): typical trajectories combine periods of high emissions with periods of no emissions \cite{Nagourney1986,Sauter1986,Bergquist1986}.

\subsection{Thermodynamics of trajectories and dynamical phase coexistence}

The statistics of emissions at long times can be obtained via the method of large deviations \cite{Garrahan2010}. If $K$ denotes the total number of emissions in a quantum trajectory up to time $t$, the probability to observe $K$ emissions (assuming perfect detection efficiency) is 
\begin{equation}
P_t(K) = \sum_{\rm traj} \delta\left( K_{\rm traj} - K \right) \approx e^{- t \varphi(K/t)} 
\end{equation}
where the sum is over the trajectories of the dynamics, and the approximate equality holds at long times where we assume that a large deviation principle holds. The function $\varphi(k)$ is the LD rate function \cite{Touchette2009}. It is related to scaled cumulant generating function (SCGF)  obtained from the moment generating function (MGF) at long times
\begin{equation}
Z_t(s) = \sum_K e^{-s K} P_t(K) \approx e^{t \theta(s)} 
\end{equation}
The SCGF $\theta(s)$ is given by the largest eigenvalue of the {\em tilted generator} \cite{Garrahan2010,Esposito2009}
\begin{equation}
\L_s(\cdot) = -i [ H, (\cdot) ] + e^{-s} J (\cdot) J^\dagger - \frac{1}{2} \{ J^\dagger J, (\cdot) \}
\label{Ls}
\end{equation}
In terms of the tilted generator, the MGF reads
\begin{equation}
Z_t(s) = \Tr e^{t \L_s} \rho_0
\label{Zs}
\end{equation}

For the 3-level model the SCGF is easy to compute by direct diagonalisation of $\L_s$. Under intermittency conditions, $\gamma,\Omega_{01} \gg \Omega_{02}$, the SCGF has the form of a free-energy displaying a (smoothed) first-order transition between two phases, one active with plentiful emissions and one inactive with no emissions \cite{Garrahan2010}. Typical dynamics occurs at coexistence between these phases, with intermittency being its dynamical manifestation. In this perspective, long periods without emissions correspond to {\em bubbles} of the inactive phase in the active one, a dynamical version of say vapour bubbles in a liquid phase. The physics is thus of a rounded first order transition (the smoothing due to the fact that the system is finite), occurring in trajectory space rather than in state space \cite{Garrahan2010}.

\section{Controlled dynamics}
\label{sec.controlled}

\subsection{Unitary control after a period without jumps and its LD properties}

We now implement a simple control scheme similar to that of the experiment of Ref.~\cite{Minev2018}.  The idea is the following. When monitoring the environment (by observing the emissions into it), a long enough survival time without emissions is an indicator of (the possibility) of a transition from a period of high emissions to one of no emissions. That is, a long survival time can serve as a warning of the system about to become shelved in the dark state. This information can be used to make an (instantaneous for simplicity) unitary perturbation which should help {\em reverse the jump}. 

While analysing such a controlled dynamics in a general system is not simple - as in principle the Markovian character is lost by the conditioned control - for the three level model considered here there is an important simplification. The original dynamics in this case that of a (quantum) {\em renewal process}, that is, after each emission the state of the system is reverted to $| 0 \rangle \langle 0 |$. And since the condition to trigger an action depends only on the time $\Delta t$ survived after since the last jump, the controlled dynamics is also that of a renewal process. This is the property that will allow us to solve the controlled dynamics in relatively simple terms. 

In the original (uncontrolled) dynamics, the state evolves between times $t_1$ and $t_2$ with no emissions according to the evolution super-operator, cf.\ Sec.\ II, 
\begin{equation}
\Q_{t_2,t_1} = e^{(t_2-t_1) \R} ,
\label{Q}
\end{equation}
generated by 
\begin{equation}
\R(\cdot) = -i [ H, (\cdot) ] - \frac{1}{2} \{ J^\dagger J, (\cdot) \}
\label{R}
\end{equation}
A state initially in $\rho_{t_1}$ conditioned on no emissions occurring up to time ${t_2}$ therefore reads
\begin{equation}
\Q_{t_2,t_1} \rho_{t_1} = e^{(t_2-t_1) \R} \rho_{t_1}
\end{equation}
Note that the resulting state above is not normalised, as the normalisation, $\Tr \Q_{t_1,t_2} \rho_{t_1}$, is the probability of the no-jump condition over that period.

In order to implement the control, we define a new dynamics, where the no-jump evolution is now given by 
\begin{equation}
\Q_{t_2,t_1}^U
= \left\{ 
\begin{array}{ccc}
e^{(t_2 - t_1) \R}  & & (t_2-t_1) < \Delta t \\
\\
e^{(t_2-t_1-\Delta t) \R} \, \U \, e^{\Delta t \R}     & & (t_2-t_1) \geq \Delta t \\
\end{array}
\right.
\label{QU}
\end{equation}
where $\U$ is the super-operator 
\begin{equation}
\U(\cdot) = U (\cdot) U^\dagger
\end{equation}
corresponding to the action of the unitary $U$ on a state. In \er{QU} we assume that the initial time $t_1$ is that when a jump has just occurred. 

The above implements the control scheme: if the waiting time between events reaches a time $\Delta t$, this is taken as the indication to act on the system with unitary $U$. As in the experiment of Ref.~\cite{Minev2018}, the aim is to choose $U$ and $\Delta t$ such that the crossover to the inactive period is reversed by the action of $U$. Furthermore, we define the jump part of the dynamics to be the same as in the uncontrolled case.

\begin{figure*}[t!]
\begin{center}
\includegraphics[width=\textwidth]{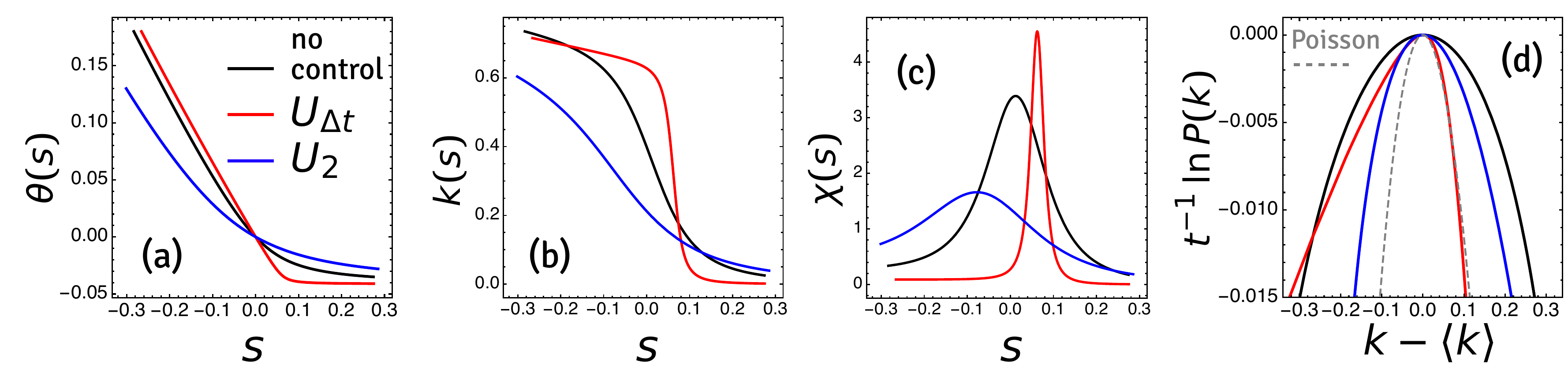}
\caption{
(a) LD scaled cumulant generating functions $\theta(s)$ 
for the original dynamics (black), for the controlled dynamics with $U_{\Delta t}$ and $\Delta t = 3$ (red), and for the controlled dynamics with $U_{2}$ and $\Delta t = 3/2$ (blue). (b) $s$-dependent activities (scaled number of emissions) $k(s) = -\theta'(s)$ for the three dynamics. (c) Activity susceptibilities (variances of $k$ scaled by time) $\chi(s) = \theta''(s)$ for the three dynamics (the red curve divided by $3$ for ease of comparison). (d) Scaled logarithm of the probability of the number of emissions in the three dynamics; abscissa shifted by $\langle k \rangle$. The grey dashed curve is the corresponding Poisson distribution with the same average as the black curve.  Parameters: $\Omega_{01}=1,\Omega_{02}=1/10,\gamma=4$.
}
\end{center}
\end{figure*}

This new dynamics cannot be generated as that of \ers{V}{L}. We can still write the new evolution operator $\T_{t_2,t_1}^U$ in terms of the Dyson series that defines its action on the state. Consider first the Dyson series for the original dynamics. In that case the evolution operator $\T_{t,0} = e^{t \L}$ can be written as
\begin{align}
\T_{t,0} = 
\sum_{K=0}^\infty \int_{0 \leq t_1 \cdots t_K \leq t} 
\Q_{t_K,t} \, \J \, \Q_{t_K,t_{K-1}} \cdots \J \, \Q_{t_1,0}
\label{Dyson}
\end{align}
where the super-operator,
\begin{equation}
\J(\cdot) = J (\cdot) J^\dagger 
\label{JJ}
\end{equation}
is the part of $\L$ responsible for the jumps (and $\L = \J + \R$), and $\Q$ is given in \er{Q}. The sum in \er{Dyson} over the number of jumps and the integral over all possible jump times corresponds to summing over all possible quantum trajectories (we have set the initial time to $0$ and the final one to $t$). We can write a similar expression to \er{Dyson} for the controlled dynamics by replacing $\Q$ by $\Q^{U}$, cf.\ \er{QU}, 
\begin{align}
\T_{t,0}^U = 
\sum_{K=0}^\infty \int_{0 \leq t_1 \cdots t_K \leq t} 
\Q_{t_K,t}^U \, \J \, \Q_{t_K,t_{K-1}}^U \cdots \J \, \Q_{t_1,0}^U
\label{DysonU}
\end{align}

While, in contrast to the original evolution operator, there is no generator for $\T_{t_1,t_2}^U$ we can still tilt the controlled evolution operator, $\T_{t_1,t_2}^U \to \T_{t_1,t_2}^{U,s}$ by making the change in \er{DysonU}, $\J \to \J_s = e^{-s} \J$. The corresponding MGF then reads, 
\begin{equation}
Z_t^U(s) = \Tr \T_{t,0}^{U,s} \rho_0
\label{ZsU}
\end{equation}
Given that the dynamics due to $\T_{t_1,t_2}^U$ will have finite correlation times despite the control, we still expect the MGF to have a LD form at long times,
\begin{equation}
Z_t^U(s) \approx e^{t \theta_U(s)}
\label{thetaU}
\end{equation}

To obtain $\theta_U(s)$ it will prove convenient to Laplace transform in time the Dyson series for the tilted evolution operator [given by \er{DysonU} with $\T_{t,0}^{U}$ replaced by $\T_{t,0}^{U,s}$ and all $\J$ replaced by $\J_s$]. The Laplace transformation makes all the time convolutions into products,
\begin{align}
\hT_{x}^{U,s} & =  
\sum_{K=0}^\infty \hQ_{x}^U \, \J_s \, \hQ_{x}^U \cdots \J \, \hQ_{x}^U 
\nonumber \\
& = \hQ_{x}^U \left( \I - \J_s \, \hQ_{x}^U \right)^{-1}
\label{Dysonx}
\end{align}
where $\hat{f}_x = \int_{0}^\infty dt e^{-x t} f_t$, $\I$ is the super-operator identity, and we have used the fact that 
$\Q_{t_1,t_2}^U$ only depends on the time difference, cf.~\er{QU}. 

We can write $\hT_{x}^{U,s}$ as
\begin{align}
\hT_{x}^{U,s} & =  \hQ_{x}^U \left( \I - e^{-s} \F_x^U \right)^{-1}
\label{DysonF}
\end{align}
where from cf.~\er{QU} we can write $\hQ_{x}^U$ as 
\begin{equation}
\hQ_{x}^U = \left( x \, \I - \R \right)^{-1} \left[ \I - e^{-\Delta t \, x} \left( \I - \U \right) e^{\Delta t \R} \right]
\label{hQU}
\end{equation}

The super-operator
\begin{align}
\F_x^U = \J \left( x \, \I - \R \right)^{-1} \left[ \I - e^{-\Delta t \, x} \left( \I - \U \right) e^{\Delta t \R} \right]
\label{FU}
\end{align}
is the tilted evolution super-operator in the so-called $x$-ensemble of trajectories of fixed number of jumps but fluctuating total overall time \cite{Budini2014,Kiukas2015}. That is, at $x=0$, $\F^U = \F_{X=0}^U$ is the evolution operator that evolves the average state between two jumps, irrespective of the time elapsed between jumps. It is easy to check from \er{FU} that $\F^U$ is probability conserving, $\Tr \F^U \rho=\Tr \rho$. The case $x \neq 0$ corresponds to a tilting of $\F^U$ associated to the SCGF for the statistics of the total time in trajectories with a fixed number of jumps \cite{Budini2014,Kiukas2015}. 

\Ers{DysonF}{FU} express the tilted evolution operator of the controlled dynamics in terms of the original dynamics, the unitary $U$ and time $\Delta t$. Having an explicit form for $\F_x^U$ allows to obtain the SCGF for the controlled process, cf.\ \er{thetaU}, in a manner that generalises the procedure of Ref.~\cite{Budini2014}. 

The action ``to the left'' of super-operators such as \era{Dysonx}{FU} is defined as the action of their adjoints on operators \cite{Breuer2002,Gardiner2004}. That is, if $\S$ is a super-operator and $X$ an operator, then the resulting operator after applying $\S$ to the left to $X$ is
\begin{equation}
X \, \S \equiv  \S^* X 
\label{left}
\end{equation}
where $\S^*$ is the adjoint of $\S$ with respect to the inner product $(X,Y)= {\rm Tr}(X^*Y)$ . Using this notation, we call $L_s$ the left-eigenoperator of $\T_{t}^{U,s}$ in \er{DysonU} corresponding to its dominant eigenvalue, which we assume has the form of the r.h.s.\ of \er{thetaU}
\begin{equation}
L_s \T_{t}^{U,s} = e^{t \theta_U(s)} L_s
\end{equation}
where $\theta_U(s)$ is what we are trying to evaluate. For the Laplace space operator \er{Dysonx} then we have
\begin{equation}
L_s \hT_{x}^{U,s} = \left[ x - \theta_U(s) \right]^{-1} L_s
\label{LLS}
\end{equation}
Using \er{Dysonx} this equation can be rearranged to 
\begin{equation}
L_s \J = L_s \left( \hQ_{x}^U \right)^{-1} - \left[ x - \theta_U(s) \right]^{-1} L_s
\label{LsQ}
\end{equation}

From the poles of \era{DysonF}{LLS} we see the relation between the SCGF $\theta_{U}(s)$ and the corresponding largest eigenvalue $e^{g_U(x)}$ of $\F_x^U$. Namely, \er{LLS} diverges when $x = \theta_{U}(s)$, while \er{LLS} does so when $g_U(x) = s$. Since $\theta_{U}(0) = 0$ and $g_{U}(0) = 0$ by probability conservation, by continuity of the largest eigenvalues we get that $g_U$ is given by the inverse function of $\theta_U$ and vice-versa \cite{Budini2014,Kiukas2015}
\begin{equation} 
g_U(x) = \theta_U^{-1} \Leftrightarrow \theta_U = g_U^{-1} 
\label{gU}
\end{equation}
where $\theta_U^{-1}[\theta_U(s)]=s$ and $g_U^{-1}[g_U(x)]=x$. This means that we can obtain the SCGF of the controlled process, \er{thetaU}, and thus all the statistical properties of the dynamics at long times, from diagonalising the operator $\F_x^U$, for which we have an explicit form, \er{FU}.

\subsection{Control and jump reversal} 

Now we test the effect of the control using the definitions above. We will use two different kinds of control operators $U$. For the first one we choose as a unitary the transformation in the $|0\rangle,|2\rangle$ subspace that rotates away the $|2\rangle$ component. That is, if after no-jump evolution up to time $\Delta t$ we have
\begin{equation}
|\psi_{\Delta t} \rangle = \frac{e^{-i \Delta t H_{\rm eff}} |0\rangle}
{\sqrt{\Vert e^{-i \Delta t H_{\rm eff}} |0\rangle \Vert^2}}
 = a \, |0 \rangle + b \, |1\rangle + c \, |2\rangle
\label{psidt}
\end{equation}
with $|a|^{2}+|b|^{2}+|c|^{2}=1$, then we choose $U_{\Delta t}$ such that 
\begin{equation}
U_{\Delta t} |\psi_{\Delta t} \rangle = a \, |0 \rangle + \sqrt{|b|^{2}+|c|^{2}} \, |1\rangle
\label{UDt}
\end{equation}
We have labelled these unitaries by ${\Delta t}$, as the form of the operator depends on the time at which it is applied given that the state to be rotated changes with time, cf.\ \er{psidt}. This control is similar to the one applied in the experiment of Ref.~\cite{Minev2018}.

The second kind of control we consider is in terms of a unitary corresponding to a $\pi/2$ rotation in the $\{ |0\rangle,|2\rangle \}$ subspace, that is, we define an operator $U_2$,
\begin{equation}
U_2 = \exp \left[ \frac{i \pi}{2} \left( | 0 \rangle \langle 2 | +
| 2 \rangle \langle 0 | \right) \right] 
\label{U2}
\end{equation}
such that its action on a state like the one of \er{psidt} is
\begin{equation}
U_2 |\psi_{\Delta t} \rangle
= i c  |0 \rangle + b |1\rangle + i a |2\rangle
\end{equation}
irrespective of the value of $\Delta t$. 

The survival function for the uncontrolled dynamics
\begin{equation}
S(t) = \Tr \Q_{t,0} |0\rangle \langle0| 
\end{equation}
corresponds to the probability of having had no jumps up to time $t$ after the last jump at time $0$. The survival probability in the controlled dynamics is 
\begin{equation}
S^U(t) = \Tr \Q_{t,0}^U |0\rangle \langle0|
\end{equation}
Figure 1(a) compares the survival probabilities without and with control. The uncontrolled case has the characteristic two-time structure when $\gamma,\Omega_{01} \gg \Omega_{02}$. For the controlled dynamics, the probability of surviving beyond $\Delta t$ changes, depending on the control operator $U$ and on the time $\Delta t$.
For control with $U_{\Delta t}$ we choose $\Delta t \, \Omega_{01} = 3$. This corresponds to a time where the occupations of the three states is comparable, see Inset to Fig.\ 1(c) (dashed line). Since the application of $U_{\Delta t}$ rotates the state to remove all projection on the dark subspace, cf.\ \er{UDt}, this control greatly reduces the likelihood of dynamics surviving beyond $\Delta t$ without an emission (but does not eliminate it completely). For control with $U_{2}$ we choose instead $\Delta t \, \Omega_{01} = 3/2$. This is a time where occupation of states $0,1$ is much higher than that of $2$ in the original dynamics, see Inset to Fig.\ 1(c) (dotted line). In this case, the action with $U$ at $\Delta t$ leads to the distinction between the two dynamical regimes to be attenuated and the survival $S(t)$ looks more like that of a single kind of dynamics that is slower. Figures 1(b) an 1(c) show representative trajectories for the three dynamics: while the original trajectories are intermittent and display dynamical coexistence, those for the two kinds of control are either entirely of high activity or entirely of low activity.

\begin{figure}[b!]
\begin{center}
\includegraphics[width=\columnwidth]{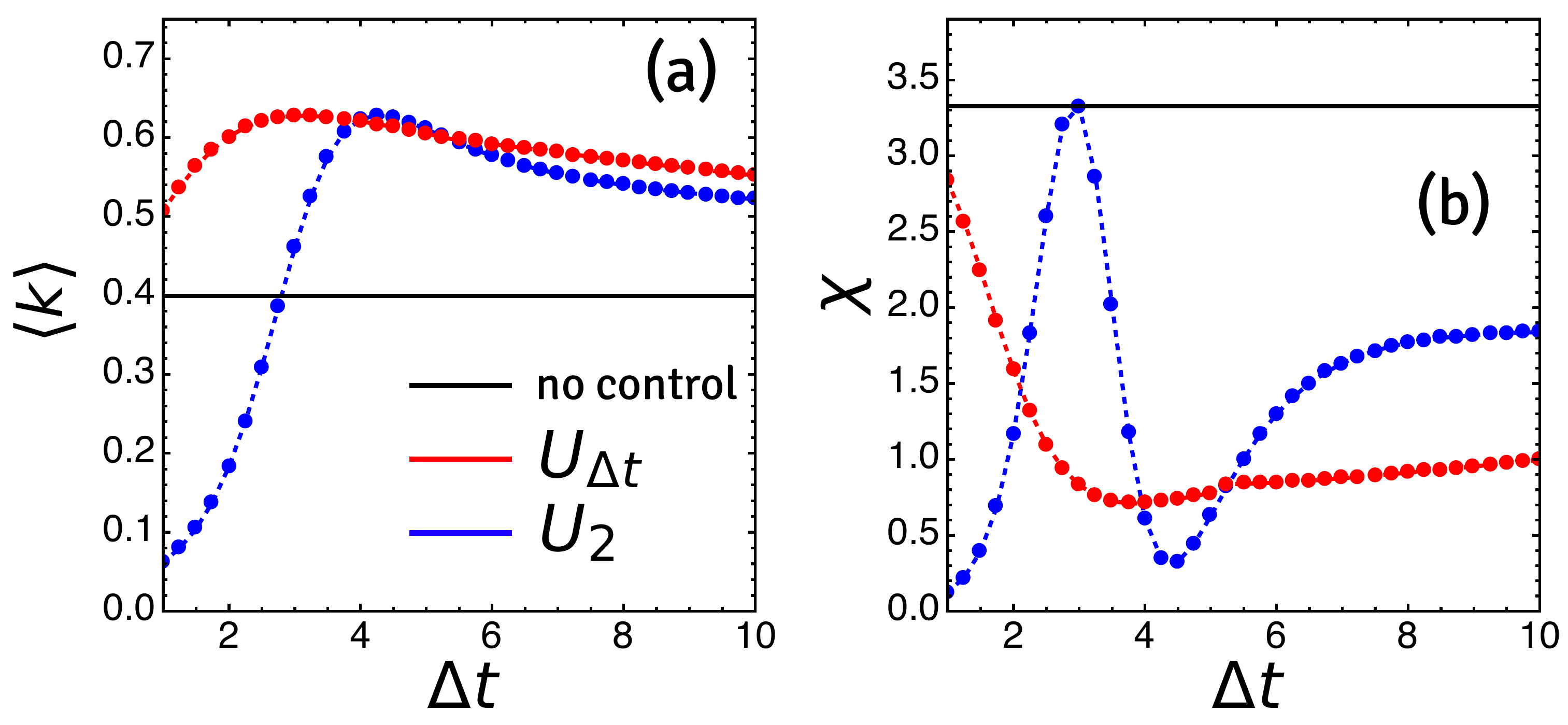}
\caption{(a) Average rate of emissions $\langle k \rangle = -\theta'(0)$ as a function of control time $\Delta t$, for the original dynamics (black - independent of $\Delta t$), for the controlled dynamics with $U_{\Delta t}$ (red), and for the controlled dynamics with $U_{2}$ (blue). (b) 
Corresponding emission susceptibilities $\chi = \theta''(0)$. Parameters: $\Omega_{01}=1,\Omega_{02}=1/10,\gamma=4$. 
}
\end{center}
\end{figure}

In Fig.\ 2 we compare the fluctuation properties of the original dynamics to the controlled dynamics, as quantified from the LD analysis above, for the same choices of $U$ and $\Delta t$. Figure 2(a) shows the SCGF $\theta(s)$
for all cases, where for the controlled dynamics $\theta(s)$ is obtained by diagonalising $\F_x^U$ to get the SCGF $g(x)$ for fixed number of jumps and then using \er{gU}.
Figure 2(b) shows the corresponding $s$-dependent number of emissions, $k(s) = -\theta'(s) = t^{-1} \langle e^{-s K} K \rangle/\langle e^{-s} \rangle$. Depending on the nature of the control we see two possible effects for our choices of $\Delta t$. The value of $s$ corresponding to the dynamical crossover - the point at which dynamics is most fluctuating - is determined by the location of the peak of the dynamical susceptibility, $\chi(s) = \theta''(s) = t^{-1} \left[ \langle e^{-s K} K^2 \rangle/\langle e^{-s K} \rangle - t^2 k^2(s) \right]$, Fig.\ 2(c). For the case of $U_{\Delta t}$ this crossover moves away from $s = 0$ as compared to the original dynamics [red versus black curves in Fig.\ 2(c)]. 
The further away a feature is from $s=0$ (which corresponds to typical dynamics) means the less probable that feature will be observed in the dynamics. This means that under $U_{\Delta t}$ dynamical coexistence is suppressed making intermittency a much rarer phenomenon in the typical trajectories of the controlled dynamics; see the typical trajectory in the middle of Fig.\ 1(b). Nevertheless, for the control with $U_{\Delta t}$ the crossover does not disappear completely. It becomes {\em sharper} than in the original dynamics, but now occurs at a larger value of $s$, see Figs.\ 2(b,c). This is due to the fact that while accessing the inactive phase is much suppressed due to the control, it is not completely eliminated, and when the system manages to switch to the dark subspace despite the control, it remains there for a long time, even if that occurrence is much rarer than in the original dynamics.

The effect of control with $U_2$, for the chosen value of $\Delta t$, is different. In this case the dynamical coexistence is strongly suppressed, as manifested by the shallower nature of $k(s)$ and its susceptibility $\chi(s)$, Figs.\ 2(b,c). The behaviour is essentially that of a single phase (there is no sharpness anywhere in $s$, in contrast to both the original dynamics and the first kind of control) with a low average number of emissions; see the typical trajectory at the bottom of Fig.\ 1(b). 

Figure 2(d) compares the probabilities of the total number of emissions, $P_t(K)$, in the different dynamics. For the original dynamics (black curve), $P_t(K)$ is very broad due to the strong correlations in the dynamics - for comparison we also show a Poisson distribution with the same average (dashed curve). Both controls (red and blue curves) make $P_t(K)$ narrower than in the original dynamics, i.e., they suppress manifestations of phase coexistence in the observed fluctuations. [Note the leftmost tail in the probability for $U_{\Delta t}$ control (red curve): this is in the effect in probability of the sharp feature in the SCGF, cf.\ Fig.\ 2(a), corresponding to a ``Maxwell construction'' associated to the (almost) phase transition that occurs in very rare trajectories of very low emissions.]

The effect of the control depends on the time $\Delta t$ at which it is applied. Figure 3 shows what occurs when $\Delta t$ is varied, by comparing the average number of emissions, $k = t^{-1} \langle K \rangle = -\theta'(0)$ (panel a) and its scaled variance, $\chi = t^{-1} {\rm var}{(K)} = \theta''(0)$ (panel b), in the original dynamics (black curves) and in the controlled dynamics with $U_{\Delta t}$ (red curves) and $U_2$ (blue curves). The control with $U_{\Delta t}$ always promotes the active dynamics (as expected as it always suppresses the state $2$), while for the control with $U_2$ which phase is favoured depends on $\Delta t$, cf.\ Inset to Fig.\ 1(a).  Nevetherless, for both controls, the fluctuations in the typical dynamics are attenuated for all $\Delta t$, Fig.\ 3(b), meaning that the control suppresses intermittency by removing the phase coexistence in the typical dynamics.

\section{Control applied repeatedly in no-jump periods} 
\label{sec.repeat}

\subsection{Feedback-control scheme and its LD properties}

In the previous sections we discussed the application of a single control pulse at time $\Delta t$ after the last emission, in analogy with the experiment of Ref.\ \cite{Minev2018}. While this may suppress the likelihood of entering the dark inactive phase, it does not suppress it completely, cf.\ Fig.\ 2: for control with $U_{\Delta t}$, at $\Delta t$ after the application of the unitary the population in $2$ is completely removed, but nothing prevents a further build up of this population in the (very unlikely) event of a subsequent long period without emissions.

One can then consider a repeated application of the control at intervals of $\Delta t$ of no emissions have occurred. The analysis of Sec.\ III can be extended for this new dynamics. Generalising \er{QU} we can define
\begin{equation}
\Q_{t_2,t_1}^{U,M}
= \left\{ 
\begin{array}{ccc}
e^{\tau \R}  & & \tau \in [0, \Delta t) \\
\\
e^{(\tau-\Delta t) \R} \, \U \, e^{\Delta t \R}     & & \tau \in [\Delta t, 2 \Delta t) \\
\vdots \\
e^{(\tau-M\Delta t) \R} \, \left( \U \, e^{\Delta t \R} \right)^M     & & \tau \in [M \Delta t, \infty) \\
\end{array}
\right.
\label{QUM}
\end{equation}
corresponding to the application of the unitary $M$ times, where $\tau= t_2-t_1$. 
After some algebra, \er{hQU} generalises to 
\begin{align}
\hQ_{x}^{U,M} = \left( x \, \I - \R \right)^{-1} 
\left[ \I - \sum_{m=1}^M e^{- m \Delta t \, x} \left( \I - \U \right)
\right. 
\nonumber \\
\left.
\phantom{\sum_{m=1}^M}
 e^{\Delta t \R} 
\left( \U \, e^{\Delta t \R} \right)^{m-1}
\right]
\label{hQUM}
\end{align}
with \er{FU} simply becoming
\begin{align}
\F_x^{U,M} = \J \hQ_{x}^{U,M}
\label{FUM}
\end{align}

In particular, we can consider the case $M \to \infty$, i.e., the unitary is applied at every interval $\Delta t$ during the no emissions period until an emission occurs. In this case \era{hQUM}{FUM} become
\begin{align}
\label{hQUMnew}
\hQ_{x}^{U,\infty} = 
\frac{ \I - e^{\Delta t (\R -x\I )} }{ x \, \I - \R } (1- \U e^{\Delta t (\R -x\I )})^{-1}
\end{align}
and
\begin{align}
\F_x^{U,\infty} = \J \hQ_{x}^{U,\infty}
\label{FUinf}
\end{align}

Proceeding like in the previous section we can obtain the fluctuation properties of the dynamics for the repeated control at long times from the largest eigenvalue of $\F_x^{U,\infty}$ via \er{gU}. Figure 4 shows the results when the unitary applied repeatedly at intervals $\Delta t$ is $U_{\Delta t}'$ defined such that, cf.\ \era{psidt}{UDt},
\begin{equation}
U_{\Delta t}' |\psi_{\Delta t} \rangle = |0 \rangle 
\label{UDt'}
\end{equation}
Both the SCGF, Fig.\ 4(a), and the $s$-dependent activity, Fig.\ 4(b), are completely smooth in $s$, and all hint of two dynamical phases has disappeared, as also confirmed by the virtual absence of any peak in the $s$-dependent susceptibility in the repeated control case, see Inset to Fig.\ 4(b). Correspondingly, the probability of the number of emissions $P_t(K)$ for the control case becomes essentially Poissonian, see Fig.\ 4(c). Figure 4(d) confirms this observations for other values of $\Delta t$. The physics is clear: the repeated action with the unitary prevents the system from populating the dark state, so that in the controlled case a single phase of the dynamics survives.

\begin{figure*}[t!]
\begin{center}
\includegraphics[width=\textwidth]{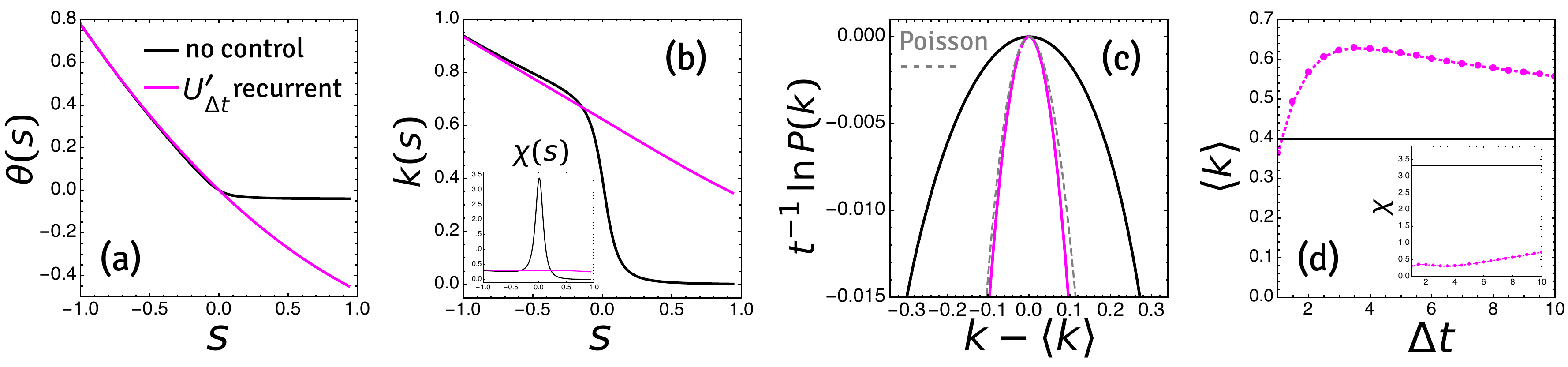}
\caption{Repeated control within no-jump periods. 
(a) LD scaled cumulant generating functions $\theta(s)$ 
for the original dynamics (black) and for the controlled dynamics $U_{\Delta t}'$ applied repeatedly at intervals $\Delta t = 3$  within no-emission periods (magenta). (b) $s$-dependent activities $k(s) = -\theta'(s)$ for the two dynamics. Inset: Activity susceptibilities $\chi(s) = \theta''(s)$ for the three dynamics. (c) Scaled logarithm of the probability of the number of emissions in the two dynamics; abscissa shifted by $\langle k \rangle$. The grey dashed curve is the corresponding Poisson distribution with the same average as the black curve.  (d) Average rate of emissions $\langle k \rangle = -\theta'(0)$ as a function of control time $\Delta t$ for the two dynamics. Inset: Corresponding emission susceptibilities $\chi = \theta''(0)$. Parameters: $\Omega_{01}=1,\Omega_{02}=1/10,\gamma=4$. 
}
\end{center}
\end{figure*}

\subsection{Markovian implementation with a classical controller}

The way we implemented the feedback dynamics in Secs.\ III and IV.A relies on applying a conditional control unitary at certain time intervals $\Delta t$. A consequence of this is that the system evolution is not described by a Markov semigroup anymore, in contrast to the original uncontrolled dynamics, cf.\ \ers{V}{L}. This prevented us from applying the usual LD approach for studying the trajectory ensemble of Markovian systems [based on the properties of a tilted generator such as \er{Ls}, the so-called $s$-ensemble method]. To circumvent this problem we exploited the so-called $x$-ensemble method \cite{Budini2014,Kiukas2015} for ensembles of trajectories with fixed number of jumps (rather than fixed overall time), as in that case the controlled dynamics can still be described by a single discrete transition operator per jump, cf.\ \era{FU}{FUinf}. 

We now show that the above control procedure can also be described in a Markovian way using  an additional classical controller which keeps track of the time elapsed from the last jump. (For other instances of a Markovian formulation of a non-Markovian dynamics via an extension of the state space see e.g.\ \cite{Breuer2004,Hush2015}.) 

For simplicity and clarity we discretise the time evolution in small time intervals $\delta t$ during which the system can either emit a photon or not, with corresponding Kraus operators \cite{Breuer2002} 
\begin{equation}
K_0^{\delta t} = e^{-i\delta t H} \sqrt{1- \delta t J^\dagger J} \, , 
\;\; K_1^{\delta t} = e^{-i\delta t H} \sqrt{\delta t } J 
\end{equation} 
so that the infinitesimal master evolution is given by
\begin{equation}
\T^{\delta t}(\cdot) = K_0 (\cdot) K_0^{\delta t \dagger} +K_1^{\delta t} (\cdot) K_1^{\delta t \dagger} 
\end{equation}

Let $n = \Delta t/ \delta t$ be the number of discrete steps in the time interval $\Delta t$, and consider a classical controller with configurations labelled $\{0,1, \dots,  n-1\}$. The classical counting measurement output of the 3-level system is fed-forward and drives the controller's dynamics as follows: Assuming that the controller is in state $l<n-1$, if no photon is detected in the next time interval $\delta t$, then the controller moves to $l+1$ (that is, the controller acts as a ``clock'' timing the duration of a no-jump period). In contrast, if a photon is detected the controller is reset back to zero. When the controller reaches the state $n-1$ (which indicates that a time $\Delta t$ has passed with no emissions) the unitary feedback operator $U$ is applied and the controller is reset to zero. 

This dynamics can be described by a discrete time transition operator on the hybrid quantum-classical system with state space $M_3 \otimes \mathbb{C}^n$ in terms of augmented Kraus operators that read \cite{Note1}
\begin{align}
\K^{\delta t}_0 & = 
\sum_{l=0}^{n-2}
K^{\delta t}_0 \otimes |l+1\rangle \langle l| \, + \, 
U K^{\delta t}_0 \otimes |0 \rangle \langle n-1 |  
\end{align}
and 
\begin{align}
\K^{\delta t}_1 &= 
\sum_{l=0}^{n-1}
K^{\delta t}_1 \otimes |0 \rangle \langle l|   
\end{align}
where $\{ | l \rangle : l=0,\ldots,n-1 \}$ represent the  states of the classical controller. (To lighten the expressions in what follows we do not label the operators and super-operators with $U$ in contrast to what we did in the previous sections.)  With these definitions, the joint system-controller evolution is then given by the transition operator
\begin{equation}
\TT^{\delta t}(\cdot) = 
\KK_0^{\delta t}(\cdot)+  \KK^{\delta t}_1(\cdot) 
\label{Kcont}
\end{equation}
where $\KK_{0,1}^{\delta t}(\cdot) = \K^{\delta t}_{0,1} \cdot \K_{0,1}^{\delta t \dagger}$. Since for our discretised dynamics \er{Kcont} is the operator that ``generates'' it, we can study the associated LD properties via the standard approach, cf.\ Sec.\ II.B, by defining the tilted operator 
\begin{equation}
\TT^{\delta t}_s(\cdot) = \KK^{\delta t}_0(\cdot)+  e^{-s}\KK^{\delta t}_1(\cdot)
\label{VVs}
\end{equation}
and finding its largest eigenvalue. 

We now confirm that this yields the same result as in Sec.\ IV.A. The $x$-ensemble formulation for trajectories with a fixed number of jumps $K$ in the system-controller dynamics has evolution operator for each jump
\begin{equation}
\FF^{\delta t}_x = \sum_{k=0}^\infty \KK^{\delta t}_1  (e^{-x\delta t} \KK^{\delta t}_0)^k 
=   \KK^{\delta t}_1 ( \II - e^{-x\delta t} \KK^{\delta t}_0 )^{-1}
\label{Fcont}
\end{equation}
corresponding to an arbitrary application of the no-jump step $\KK_0^{\delta t}$ terminated by a jump $\KK_1^{\delta t}$, and where each step is weighted by $e^{-x\delta t}$, cf.\ \era{FU}{FUinf}. Note that $\FF^{\delta t}_x$ maps any state into the product of the $0$ state of the atom and the $0$ state for the controller, $\varrho_{00} = |0\rangle \langle 0| \otimes |0\rangle \langle 0|$.
That is, if the joint state of the system-controller is
\begin{equation}
\varrho = \sum_{l=0}^{n-1} \rho_l \otimes |l \rangle \langle l| 
\label{varrho}
\end{equation}
then
\begin{equation}
 \KK^{\delta t}_1(\varrho) = \gamma \sum_{l=0}^{n-1} \langle 1  | \rho_l |1\rangle  \, \varrho_{00} 
= \left( \sum_{l=0}^{n-1} \J \rho_l \right) \otimes |0 \rangle \langle 0| 
\label{KK1}
 \end{equation}
where we have used \er{JJ} in the last equality. 

Consider now the action of the operator \er{Fcont} on the combined, $\varrho = \rho_0 \otimes | 0 \rangle \langle 0 |$, which would correspond either to the initial state or to the state just after a jump ($\rho_0 = | 0 \rangle \langle 0 |$). From \era{Fcont}{KK1} we have that
\begin{equation}
\FF^{\delta t}_x \varrho = \left( \sum_{l=0}^{n-1} \J \tau_l \right) 
\otimes |0 \rangle \langle 0| 
\label{FFF}
\end{equation}
where we have defined
\begin{align}
\sum_{l=0}^{n-1} \tau_l \otimes |l \rangle \langle l| = 
( \II - e^{-x\delta t} \KK^{\delta t}_0 )^{-1} \varrho 
\end{align}
The equation above can be inverted to give
\begin{align}
\tau_l &= \left( e^{-x \delta t} \KKK_0^{\delta t} \right)^l \left[ \I - \U \left( e^{-x \delta t} \KKK_0^{\delta t} \right) \right]^{-1} \rho_0
\label{tau}
\end{align}
where $\KKK_0^{\delta t} (\cdot) = K^{\delta t}_0 (\cdot) K_0^{\delta t \dagger}$. 
By taking the limit $\delta t$ to zero (and consequently $n = \Delta t/\delta_t$ to infinity) the sum in \er{FFF} becomes an integral, and using \er{tau} we can write
\begin{align}
\FF^{\delta t}_x \varrho 
&=
\int_0^{\Delta t} dt 
\J 
e^{t (\mathcal{R} -x)} 
\left( \I - \U e^{t (\mathcal{R} -x)} \right)^{-1} \rho_0
\otimes |0 \rangle \langle 0| 
\nonumber \\
&= 
\J \, 
\frac{ \I - e^{\Delta t (\R -x\I )} }{ x \, \I - \R } 
\left( \I - \U e^{t (\mathcal{R} -x)} \right)^{-1} \rho_0
\otimes |0 \rangle \langle 0| 
\nonumber \\
&= \left( \F_x^{U,\infty} \rho_0 \right) 
\otimes |0 \rangle \langle 0| 
\end{align}
where we have used that $\lim_{\delta t \to 0} \left( \KKK_0^{\delta t} \right)^{t/\delta t} = e^{t \R}$. The above shows that in the limit of $\delta t \to 0$ the action of the evolution operator \er{Fcont} for the system-controller combined setup is the same as that of the non-Markovian evolution operator \er{FUinf} once the controller is traced out. This implies that in this limit the SCGF function extracted from the tilted generator \er{VVs} is the same as that obtained in Sec.\ IV.A.

\section{Conclusions}

Here we have considered feedback-control schemes based on the one applied in the experiment of Ref.\ \cite{Minev2018} to catch and revert quantum jumps. (For other control schemes aiming to temper intermittency see for example \cite{Mabuchi2011,Wu2018}.)
We have focused on the case of an intermittent three level system, as in the experiment, but our approach can be generalised to more complex situations. We have considered in particular the statistics of dynamical fluctuations by means of large deviation methods. Our central observation is that unitary perturbations applied conditioned on the time elapsed without emissions lead to a modification of the observed dynamics where intermittency is either suppressed or completely removed. From the thermodynamics of trajectories perspective \cite{Garrahan2010,Lesanovsky2013}, intermittency is due to the existence of a nearby dynamical first-order transition (smoothed to a first-order crossover in a finite state system such as the one considered here) between active and inactive phases: the controlled dynamics suppresses this transition by favouring one or the other dynamical phase, depending on the details of the control operation. Within this thermodynamic of dynamics perspective, catching and reversing a quantum jump is similar to pricking the (space-time) bubbles of the inactive phase that give rise to the intermittent dynamics. 

On the technical level, we have used the tools of large deviations to obtain our results. Since the control dynamics as defined in Secs.\ III and IV.A (single, or multiple, application of a unitary perturbation during no-jump regimes, respectively) are not Markovian, we have exploited trajectory ensemble equivalence \cite{Chetrite2013,Kiukas2015} (yet another thermodynamic analogy) to recover the LD properties for trajectories of fixed total time from that of trajectories of fixed number of jumps. 
In Sec.\ IV.B we showed that the repeated control scheme of Sec.\ IV.A can be implemented in a Markovian, time-independent fashion, using classical feed-forward and feedback with an auxiliary classical controller. The single-application unitary scheme of Sec.\ III can also be implemented by further enlarging the controller such that it keeps track of the bit of information regarding whether the unitary has been already applied since the last photon emission. This opens the way for treating more complicated system dynamics, with controllers which (partly) encode the measurement trajectory and trigger the control action in a Markovian fashion. In particular, an interesting type of controller would be based on sliding window encoder of a finite portion of the trajectory. Here one can use ideas from symbolic dynamics and coding \cite{LindMarcus}. More generally, the controller can be fully quantum, rather than classical, exploiting the methods of coherent feedback and quantum networks theory \cite{Gough2009}. 

In future work we hope to generalise the control schemes described above to more complex setting involving multiple jump operators, multiple dynamical phases and many-body models. Furthermore, the control approach we studied here, based on that used in Ref.\ \cite{Minev2018}, in turn can lead to new thinking in the study of complex non-equilibrium dynamics. In a system with interesting rare dynamical fluctuations, as revealed by LD methods, a general problem of interest is how to engineer the dynamics to make those fluctuations typical. A common approach is that of the so-called generalised Doob transformation \cite{Jack2010,Garrahan2010,Chetrite2015,Garrahan2016,Carollo2018}. This can be formulated as an optimal control problem \cite{Jack2015b,Chetrite2015b} which aims to find a continuously controlled system that evolves (in the long time limit) with a new dynamics that is time homogeneous if the original one was so. The Doob transformed dynamics can thus be defined to target atypical dynamics of the original problem, for example at some non-zero value of the counting field $s$ in order to stabilise a dynamical phase. An interesting question is therefore to what extent one can use instead a control scheme that is local in time applied conditionally on the observed dynamics, similarly to what was done above. The time-local control might offer an efficient alternative to the Doob transform \cite{Jack2010,Garrahan2010,Chetrite2015,Garrahan2016,Carollo2018}
for generating interesting trajectory ensembles in a manner reminiscent of classical stabilisation via resets \cite{evans2011,evans2014,reuveni2016,pal2017}.

\acknowledgments
The work described here was started at the Institut Henri Poincar\'e during its programme ``Measurement and control of quantum systems: theory and experiment''. This work was supported by EPSRC Grant No.~EP/M014266/1.

\bibliography{catch}

\begin{thebibliography}{67}%
\makeatletter
\providecommand \@ifxundefined [1]{%
 \@ifx{#1\undefined}
}%
\providecommand \@ifnum [1]{%
 \ifnum #1\expandafter \@firstoftwo
 \else \expandafter \@secondoftwo
 \fi
}%
\providecommand \@ifx [1]{%
 \ifx #1\expandafter \@firstoftwo
 \else \expandafter \@secondoftwo
 \fi
}%
\providecommand \natexlab [1]{#1}%
\providecommand \enquote  [1]{``#1''}%
\providecommand \bibnamefont  [1]{#1}%
\providecommand \bibfnamefont [1]{#1}%
\providecommand \citenamefont [1]{#1}%
\providecommand \href@noop [0]{\@secondoftwo}%
\providecommand \href [0]{\begingroup \@sanitize@url \@href}%
\providecommand \@href[1]{\@@startlink{#1}\@@href}%
\providecommand \@@href[1]{\endgroup#1\@@endlink}%
\providecommand \@sanitize@url [0]{\catcode `\\12\catcode `\$12\catcode
  `\&12\catcode `\#12\catcode `\^12\catcode `\_12\catcode `\%12\relax}%
\providecommand \@@startlink[1]{}%
\providecommand \@@endlink[0]{}%
\providecommand \url  [0]{\begingroup\@sanitize@url \@url }%
\providecommand \@url [1]{\endgroup\@href {#1}{\urlprefix }}%
\providecommand \urlprefix  [0]{URL }%
\providecommand \Eprint [0]{\href }%
\providecommand \doibase [0]{http://dx.doi.org/}%
\providecommand \selectlanguage [0]{\@gobble}%
\providecommand \bibinfo  [0]{\@secondoftwo}%
\providecommand \bibfield  [0]{\@secondoftwo}%
\providecommand \translation [1]{[#1]}%
\providecommand \BibitemOpen [0]{}%
\providecommand \bibitemStop [0]{}%
\providecommand \bibitemNoStop [0]{.\EOS\space}%
\providecommand \EOS [0]{\spacefactor3000\relax}%
\providecommand \BibitemShut  [1]{\csname bibitem#1\endcsname}%
\let\auto@bib@innerbib\@empty
\bibitem [{\citenamefont {Plenio}\ and\ \citenamefont
  {Knight}(1998)}]{Plenio1998}%
  \BibitemOpen
  \bibfield  {author} {\bibinfo {author} {\bibfnamefont {M.~B.}\ \bibnamefont
  {Plenio}}\ and\ \bibinfo {author} {\bibfnamefont {P.~L.}\ \bibnamefont
  {Knight}},\ }\href@noop {} {\bibfield  {journal} {\bibinfo  {journal} {Rev.
  Mod. Phys.}\ }\textbf {\bibinfo {volume} {70}},\ \bibinfo {pages} {101}
  (\bibinfo {year} {1998})}\BibitemShut {NoStop}%
\bibitem [{\citenamefont {Belavkin}(1999)}]{Belavkin1999}%
  \BibitemOpen
  \bibfield  {author} {\bibinfo {author} {\bibfnamefont {V.}~\bibnamefont
  {Belavkin}},\ }\href@noop {} {\bibfield  {journal} {\bibinfo  {journal} {Rep.
  Math. Phys.}\ }\textbf {\bibinfo {volume} {43}},\ \bibinfo {pages} {A405}
  (\bibinfo {year} {1999})}\BibitemShut {NoStop}%
\bibitem [{\citenamefont {Breuer}\ and\ \citenamefont
  {Petruccione}(2002)}]{Breuer2002}%
  \BibitemOpen
  \bibfield  {author} {\bibinfo {author} {\bibfnamefont {H.-P.}\ \bibnamefont
  {Breuer}}\ and\ \bibinfo {author} {\bibfnamefont {F.}~\bibnamefont
  {Petruccione}},\ }\href@noop {} {\emph {\bibinfo {title} {The theory of open
  quantum systems}}}\ (\bibinfo  {publisher} {Oxford University Press},\
  \bibinfo {year} {2002})\BibitemShut {NoStop}%
\bibitem [{\citenamefont {Gardiner}\ and\ \citenamefont
  {Zoller}(2004)}]{Gardiner2004}%
  \BibitemOpen
  \bibfield  {author} {\bibinfo {author} {\bibfnamefont {C.}~\bibnamefont
  {Gardiner}}\ and\ \bibinfo {author} {\bibfnamefont {P.}~\bibnamefont
  {Zoller}},\ }\href@noop {} {\emph {\bibinfo {title} {Quantum noise}}}\
  (\bibinfo  {publisher} {Springer},\ \bibinfo {year} {2004})\BibitemShut
  {NoStop}%
\bibitem [{\citenamefont {Minev}\ \emph {et~al.}(2018)\citenamefont {Minev},
  \citenamefont {Mundhada}, \citenamefont {Shankar}, \citenamefont {Reinhold},
  \citenamefont {Guti{\'e}rrez-J{\'a}uregui}, \citenamefont {Schoelkopf},
  \citenamefont {Mirrahimi}, \citenamefont {Carmichael},\ and\ \citenamefont
  {Devoret}}]{Minev2018}%
  \BibitemOpen
  \bibfield  {author} {\bibinfo {author} {\bibfnamefont {Z.}~\bibnamefont
  {Minev}}, \bibinfo {author} {\bibfnamefont {S.}~\bibnamefont {Mundhada}},
  \bibinfo {author} {\bibfnamefont {S.}~\bibnamefont {Shankar}}, \bibinfo
  {author} {\bibfnamefont {P.}~\bibnamefont {Reinhold}}, \bibinfo {author}
  {\bibfnamefont {R.}~\bibnamefont {Guti{\'e}rrez-J{\'a}uregui}}, \bibinfo
  {author} {\bibfnamefont {R.}~\bibnamefont {Schoelkopf}}, \bibinfo {author}
  {\bibfnamefont {M.}~\bibnamefont {Mirrahimi}}, \bibinfo {author}
  {\bibfnamefont {H.}~\bibnamefont {Carmichael}}, \ and\ \bibinfo {author}
  {\bibfnamefont {M.}~\bibnamefont {Devoret}},\ }\href@noop {} {\enquote
  {\bibinfo {title} {To catch and reverse a quantum jump mid-flight},}\
  }\bibinfo {howpublished} {arXiv:1803.00545} (\bibinfo {year}
  {2018})\BibitemShut {NoStop}%
\bibitem [{\citenamefont {Murch}\ \emph {et~al.}(2013)\citenamefont {Murch},
  \citenamefont {Weber}, \citenamefont {Macklin},\ and\ \citenamefont
  {Siddiqi}}]{Murch2013}%
  \BibitemOpen
  \bibfield  {author} {\bibinfo {author} {\bibfnamefont {K.~W.}\ \bibnamefont
  {Murch}}, \bibinfo {author} {\bibfnamefont {S.~J.}\ \bibnamefont {Weber}},
  \bibinfo {author} {\bibfnamefont {C.}~\bibnamefont {Macklin}}, \ and\
  \bibinfo {author} {\bibfnamefont {I.}~\bibnamefont {Siddiqi}},\ }\href
  {http://dx.doi.org/10.1038/nature12539} {\bibfield  {journal} {\bibinfo
  {journal} {Nature}\ }\textbf {\bibinfo {volume} {502}},\ \bibinfo {pages}
  {211 EP } (\bibinfo {year} {2013})}\BibitemShut {NoStop}%
\bibitem [{\citenamefont {Weber}\ \emph {et~al.}(2014)\citenamefont {Weber},
  \citenamefont {Chantasri}, \citenamefont {Dressel}, \citenamefont {Jordan},
  \citenamefont {Murch},\ and\ \citenamefont {Siddiqi}}]{Weber2014}%
  \BibitemOpen
  \bibfield  {author} {\bibinfo {author} {\bibfnamefont {S.~J.}\ \bibnamefont
  {Weber}}, \bibinfo {author} {\bibfnamefont {A.}~\bibnamefont {Chantasri}},
  \bibinfo {author} {\bibfnamefont {J.}~\bibnamefont {Dressel}}, \bibinfo
  {author} {\bibfnamefont {A.~N.}\ \bibnamefont {Jordan}}, \bibinfo {author}
  {\bibfnamefont {K.~W.}\ \bibnamefont {Murch}}, \ and\ \bibinfo {author}
  {\bibfnamefont {I.}~\bibnamefont {Siddiqi}},\ }\href
  {http://dx.doi.org/10.1038/nature13559} {\bibfield  {journal} {\bibinfo
  {journal} {Nature}\ }\textbf {\bibinfo {volume} {511}},\ \bibinfo {pages}
  {570 EP } (\bibinfo {year} {2014})}\BibitemShut {NoStop}%
\bibitem [{\citenamefont {Tan}\ \emph {et~al.}(2015)\citenamefont {Tan},
  \citenamefont {Weber}, \citenamefont {Siddiqi}, \citenamefont {M\o{}lmer},\
  and\ \citenamefont {Murch}}]{Tan2015}%
  \BibitemOpen
  \bibfield  {author} {\bibinfo {author} {\bibfnamefont {D.}~\bibnamefont
  {Tan}}, \bibinfo {author} {\bibfnamefont {S.~J.}\ \bibnamefont {Weber}},
  \bibinfo {author} {\bibfnamefont {I.}~\bibnamefont {Siddiqi}}, \bibinfo
  {author} {\bibfnamefont {K.}~\bibnamefont {M\o{}lmer}}, \ and\ \bibinfo
  {author} {\bibfnamefont {K.~W.}\ \bibnamefont {Murch}},\ }\href@noop {}
  {\bibfield  {journal} {\bibinfo  {journal} {Phys. Rev. Lett.}\ }\textbf
  {\bibinfo {volume} {114}},\ \bibinfo {pages} {090403} (\bibinfo {year}
  {2015})}\BibitemShut {NoStop}%
\bibitem [{\citenamefont {Foroozani}\ \emph {et~al.}(2016)\citenamefont
  {Foroozani}, \citenamefont {Naghiloo}, \citenamefont {Tan}, \citenamefont
  {M\o{}lmer},\ and\ \citenamefont {Murch}}]{Foroozani2016}%
  \BibitemOpen
  \bibfield  {author} {\bibinfo {author} {\bibfnamefont {N.}~\bibnamefont
  {Foroozani}}, \bibinfo {author} {\bibfnamefont {M.}~\bibnamefont {Naghiloo}},
  \bibinfo {author} {\bibfnamefont {D.}~\bibnamefont {Tan}}, \bibinfo {author}
  {\bibfnamefont {K.}~\bibnamefont {M\o{}lmer}}, \ and\ \bibinfo {author}
  {\bibfnamefont {K.~W.}\ \bibnamefont {Murch}},\ }\href@noop {} {\bibfield
  {journal} {\bibinfo  {journal} {Phys. Rev. Lett.}\ }\textbf {\bibinfo
  {volume} {116}},\ \bibinfo {pages} {110401} (\bibinfo {year}
  {2016})}\BibitemShut {NoStop}%
\bibitem [{\citenamefont {Belavkin}(1990)}]{Belavkin1990}%
  \BibitemOpen
  \bibfield  {author} {\bibinfo {author} {\bibfnamefont {V.}~\bibnamefont
  {Belavkin}},\ }\href@noop {} {\bibfield  {journal} {\bibinfo  {journal}
  {Lett. Math. Phys.}\ }\textbf {\bibinfo {volume} {20}},\ \bibinfo {pages}
  {85} (\bibinfo {year} {1990})}\BibitemShut {NoStop}%
\bibitem [{\citenamefont {Dalibard}\ \emph {et~al.}(1992)\citenamefont
  {Dalibard}, \citenamefont {Castin},\ and\ \citenamefont
  {M{\o}lmer}}]{Dalibard1992}%
  \BibitemOpen
  \bibfield  {author} {\bibinfo {author} {\bibfnamefont {J.}~\bibnamefont
  {Dalibard}}, \bibinfo {author} {\bibfnamefont {Y.}~\bibnamefont {Castin}}, \
  and\ \bibinfo {author} {\bibfnamefont {K.}~\bibnamefont {M{\o}lmer}},\
  }\href@noop {} {\bibfield  {journal} {\bibinfo  {journal} {Phys. Rev. Lett.}\
  }\textbf {\bibinfo {volume} {68}},\ \bibinfo {pages} {580} (\bibinfo {year}
  {1992})}\BibitemShut {NoStop}%
\bibitem [{\citenamefont {Gardiner}\ \emph {et~al.}(1992)\citenamefont
  {Gardiner}, \citenamefont {Parkins},\ and\ \citenamefont
  {Zoller}}]{Gardiner1992}%
  \BibitemOpen
  \bibfield  {author} {\bibinfo {author} {\bibfnamefont {C.}~\bibnamefont
  {Gardiner}}, \bibinfo {author} {\bibfnamefont {A.}~\bibnamefont {Parkins}}, \
  and\ \bibinfo {author} {\bibfnamefont {P.}~\bibnamefont {Zoller}},\
  }\href@noop {} {\bibfield  {journal} {\bibinfo  {journal} {Phys. Rev. A}\
  }\textbf {\bibinfo {volume} {46}},\ \bibinfo {pages} {4363} (\bibinfo {year}
  {1992})}\BibitemShut {NoStop}%
\bibitem [{\citenamefont {Carmichael}(1993)}]{Carmichael1993}%
  \BibitemOpen
  \bibfield  {author} {\bibinfo {author} {\bibfnamefont {H.}~\bibnamefont
  {Carmichael}},\ }\href@noop {} {\emph {\bibinfo {title} {An open systems
  approach to quantum optics}}}\ (\bibinfo  {publisher} {Springer, Berlin},\
  \bibinfo {year} {1993})\BibitemShut {NoStop}%
\bibitem [{\citenamefont {Mabuchi}\ and\ \citenamefont
  {Zoller}(1996)}]{Mabuchi1996}%
  \BibitemOpen
  \bibfield  {author} {\bibinfo {author} {\bibfnamefont {H.}~\bibnamefont
  {Mabuchi}}\ and\ \bibinfo {author} {\bibfnamefont {P.}~\bibnamefont
  {Zoller}},\ }\href@noop {} {\bibfield  {journal} {\bibinfo  {journal} {Phys.
  Rev. Lett.}\ }\textbf {\bibinfo {volume} {76}},\ \bibinfo {pages} {3108}
  (\bibinfo {year} {1996})}\BibitemShut {NoStop}%
\bibitem [{\citenamefont {Wiseman}(2002)}]{Wiseman2002}%
  \BibitemOpen
  \bibfield  {author} {\bibinfo {author} {\bibfnamefont {H.~M.}\ \bibnamefont
  {Wiseman}},\ }\href@noop {} {\bibfield  {journal} {\bibinfo  {journal} {Phys.
  Rev. A}\ }\textbf {\bibinfo {volume} {65}},\ \bibinfo {pages} {032111}
  (\bibinfo {year} {2002})}\BibitemShut {NoStop}%
\bibitem [{\citenamefont {van Handel}\ and\ \citenamefont
  {Mabuchi}(2005)}]{Handel2005}%
  \BibitemOpen
  \bibfield  {author} {\bibinfo {author} {\bibfnamefont {R.}~\bibnamefont {van
  Handel}}\ and\ \bibinfo {author} {\bibfnamefont {H.}~\bibnamefont
  {Mabuchi}},\ }\href@noop {} {\bibfield  {journal} {\bibinfo  {journal}
  {arXiv:quant-ph/0511221}\ } (\bibinfo {year} {2005})}\BibitemShut {NoStop}%
\bibitem [{\citenamefont {Gambetta}\ \emph {et~al.}(2008)\citenamefont
  {Gambetta}, \citenamefont {Blais}, \citenamefont {Boissonneault},
  \citenamefont {Houck}, \citenamefont {Schuster},\ and\ \citenamefont
  {Girvin}}]{Gambetta2008}%
  \BibitemOpen
  \bibfield  {author} {\bibinfo {author} {\bibfnamefont {J.}~\bibnamefont
  {Gambetta}}, \bibinfo {author} {\bibfnamefont {A.}~\bibnamefont {Blais}},
  \bibinfo {author} {\bibfnamefont {M.}~\bibnamefont {Boissonneault}}, \bibinfo
  {author} {\bibfnamefont {A.~A.}\ \bibnamefont {Houck}}, \bibinfo {author}
  {\bibfnamefont {D.}~\bibnamefont {Schuster}}, \ and\ \bibinfo {author}
  {\bibfnamefont {S.~M.}\ \bibnamefont {Girvin}},\ }\href@noop {} {\bibfield
  {journal} {\bibinfo  {journal} {Phys. Rev. A}\ }\textbf {\bibinfo {volume}
  {77}},\ \bibinfo {pages} {012112} (\bibinfo {year} {2008})}\BibitemShut
  {NoStop}%
\bibitem [{\citenamefont {Wiseman}\ and\ \citenamefont
  {Milburn}(2009)}]{Wiseman2009}%
  \BibitemOpen
  \bibfield  {author} {\bibinfo {author} {\bibfnamefont {H.~M.}\ \bibnamefont
  {Wiseman}}\ and\ \bibinfo {author} {\bibfnamefont {G.~J.}\ \bibnamefont
  {Milburn}},\ }\href@noop {} {\emph {\bibinfo {title} {Quantum measurement and
  control}}}\ (\bibinfo  {publisher} {Cambridge university press},\ \bibinfo
  {year} {2009})\BibitemShut {NoStop}%
\bibitem [{\citenamefont {Gammelmark}\ \emph {et~al.}(2013)\citenamefont
  {Gammelmark}, \citenamefont {Julsgaard},\ and\ \citenamefont
  {M\o{}lmer}}]{Gammelmark2013}%
  \BibitemOpen
  \bibfield  {author} {\bibinfo {author} {\bibfnamefont {S.}~\bibnamefont
  {Gammelmark}}, \bibinfo {author} {\bibfnamefont {B.}~\bibnamefont
  {Julsgaard}}, \ and\ \bibinfo {author} {\bibfnamefont {K.}~\bibnamefont
  {M\o{}lmer}},\ }\href@noop {} {\bibfield  {journal} {\bibinfo  {journal}
  {Phys. Rev. Lett.}\ }\textbf {\bibinfo {volume} {111}},\ \bibinfo {pages}
  {160401} (\bibinfo {year} {2013})}\BibitemShut {NoStop}%
\bibitem [{\citenamefont {Guevara}\ and\ \citenamefont
  {Wiseman}(2015)}]{Guevara2015}%
  \BibitemOpen
  \bibfield  {author} {\bibinfo {author} {\bibfnamefont {I.}~\bibnamefont
  {Guevara}}\ and\ \bibinfo {author} {\bibfnamefont {H.}~\bibnamefont
  {Wiseman}},\ }\href@noop {} {\bibfield  {journal} {\bibinfo  {journal} {Phys.
  Rev. Lett.}\ }\textbf {\bibinfo {volume} {115}},\ \bibinfo {pages} {180407}
  (\bibinfo {year} {2015})}\BibitemShut {NoStop}%
\bibitem [{\citenamefont {Garrahan}\ and\ \citenamefont
  {Lesanovsky}(2010)}]{Garrahan2010}%
  \BibitemOpen
  \bibfield  {author} {\bibinfo {author} {\bibfnamefont {J.~P.}\ \bibnamefont
  {Garrahan}}\ and\ \bibinfo {author} {\bibfnamefont {I.}~\bibnamefont
  {Lesanovsky}},\ }\href@noop {} {\bibfield  {journal} {\bibinfo  {journal}
  {Phys. Rev. Lett.}\ }\textbf {\bibinfo {volume} {104}},\ \bibinfo {pages}
  {160601} (\bibinfo {year} {2010})}\BibitemShut {NoStop}%
\bibitem [{\citenamefont {Eckmann}\ and\ \citenamefont
  {Ruelle}(1985)}]{Eckmann1985}%
  \BibitemOpen
  \bibfield  {author} {\bibinfo {author} {\bibfnamefont {J.~P.}\ \bibnamefont
  {Eckmann}}\ and\ \bibinfo {author} {\bibfnamefont {D.}~\bibnamefont
  {Ruelle}},\ }\href@noop {} {\bibfield  {journal} {\bibinfo  {journal} {Rev.
  Mod. Phys.}\ }\textbf {\bibinfo {volume} {57}},\ \bibinfo {pages} {617}
  (\bibinfo {year} {1985})}\BibitemShut {NoStop}%
\bibitem [{\citenamefont {Touchette}(2009)}]{Touchette2009}%
  \BibitemOpen
  \bibfield  {author} {\bibinfo {author} {\bibfnamefont {H.}~\bibnamefont
  {Touchette}},\ }\href@noop {} {\bibfield  {journal} {\bibinfo  {journal}
  {Phys. Rep.}\ }\textbf {\bibinfo {volume} {478}},\ \bibinfo {pages} {1}
  (\bibinfo {year} {2009})}\BibitemShut {NoStop}%
\bibitem [{\citenamefont {Esposito}\ \emph {et~al.}(2009)\citenamefont
  {Esposito}, \citenamefont {Harbola},\ and\ \citenamefont
  {Mukamel}}]{Esposito2009}%
  \BibitemOpen
  \bibfield  {author} {\bibinfo {author} {\bibfnamefont {M.}~\bibnamefont
  {Esposito}}, \bibinfo {author} {\bibfnamefont {U.}~\bibnamefont {Harbola}}, \
  and\ \bibinfo {author} {\bibfnamefont {S.}~\bibnamefont {Mukamel}},\
  }\href@noop {} {\bibfield  {journal} {\bibinfo  {journal} {Rev. Mod. Phys.}\
  }\textbf {\bibinfo {volume} {81}},\ \bibinfo {pages} {1665} (\bibinfo {year}
  {2009})}\BibitemShut {NoStop}%
\bibitem [{\citenamefont {Gaspard}(2005)}]{Gaspard2005}%
  \BibitemOpen
  \bibfield  {author} {\bibinfo {author} {\bibfnamefont {P.}~\bibnamefont
  {Gaspard}},\ }\href@noop {} {\emph {\bibinfo {title} {Chaos, Scattering and
  Statistical Mechanics}}}\ (\bibinfo  {publisher} {Cambridge University
  Press},\ \bibinfo {year} {2005})\BibitemShut {NoStop}%
\bibitem [{\citenamefont {Merolle}\ \emph {et~al.}(2005)\citenamefont
  {Merolle}, \citenamefont {Garrahan},\ and\ \citenamefont
  {Chandler}}]{Merolle2005}%
  \BibitemOpen
  \bibfield  {author} {\bibinfo {author} {\bibfnamefont {M.}~\bibnamefont
  {Merolle}}, \bibinfo {author} {\bibfnamefont {J.~P.}\ \bibnamefont
  {Garrahan}}, \ and\ \bibinfo {author} {\bibfnamefont {D.}~\bibnamefont
  {Chandler}},\ }\href@noop {} {\bibfield  {journal} {\bibinfo  {journal}
  {Proc. Natl. Acad. Sci. USA}\ }\textbf {\bibinfo {volume} {102}},\ \bibinfo
  {pages} {10837} (\bibinfo {year} {2005})}\BibitemShut {NoStop}%
\bibitem [{\citenamefont {Garrahan}\ \emph {et~al.}(2007)\citenamefont
  {Garrahan}, \citenamefont {Jack}, \citenamefont {Lecomte}, \citenamefont
  {Pitard}, \citenamefont {van Duijvendijk},\ and\ \citenamefont {van
  Wijland}}]{Garrahan2007}%
  \BibitemOpen
  \bibfield  {author} {\bibinfo {author} {\bibfnamefont {J.~P.}\ \bibnamefont
  {Garrahan}}, \bibinfo {author} {\bibfnamefont {R.~L.}\ \bibnamefont {Jack}},
  \bibinfo {author} {\bibfnamefont {V.}~\bibnamefont {Lecomte}}, \bibinfo
  {author} {\bibfnamefont {E.}~\bibnamefont {Pitard}}, \bibinfo {author}
  {\bibfnamefont {K.}~\bibnamefont {van Duijvendijk}}, \ and\ \bibinfo {author}
  {\bibfnamefont {F.}~\bibnamefont {van Wijland}},\ }\href@noop {} {\bibfield
  {journal} {\bibinfo  {journal} {Phys. Rev. Lett.}\ }\textbf {\bibinfo
  {volume} {98}},\ \bibinfo {pages} {195702} (\bibinfo {year}
  {2007})}\BibitemShut {NoStop}%
\bibitem [{\citenamefont {Hedges}\ \emph {et~al.}(2009)\citenamefont {Hedges},
  \citenamefont {Jack}, \citenamefont {Garrahan},\ and\ \citenamefont
  {Chandler}}]{Hedges2009}%
  \BibitemOpen
  \bibfield  {author} {\bibinfo {author} {\bibfnamefont {L.~O.}\ \bibnamefont
  {Hedges}}, \bibinfo {author} {\bibfnamefont {R.~L.}\ \bibnamefont {Jack}},
  \bibinfo {author} {\bibfnamefont {J.~P.}\ \bibnamefont {Garrahan}}, \ and\
  \bibinfo {author} {\bibfnamefont {D.}~\bibnamefont {Chandler}},\ }\href@noop
  {} {\bibfield  {journal} {\bibinfo  {journal} {Science}\ }\textbf {\bibinfo
  {volume} {323}},\ \bibinfo {pages} {1309} (\bibinfo {year}
  {2009})}\BibitemShut {NoStop}%
\bibitem [{\citenamefont {Speck}\ \emph {et~al.}(2012)\citenamefont {Speck},
  \citenamefont {Malins},\ and\ \citenamefont {Royall}}]{Speck2012b}%
  \BibitemOpen
  \bibfield  {author} {\bibinfo {author} {\bibfnamefont {T.}~\bibnamefont
  {Speck}}, \bibinfo {author} {\bibfnamefont {A.}~\bibnamefont {Malins}}, \
  and\ \bibinfo {author} {\bibfnamefont {C.~P.}\ \bibnamefont {Royall}},\
  }\href@noop {} {\bibfield  {journal} {\bibinfo  {journal} {Phys. Rev. Lett.}\
  }\textbf {\bibinfo {volume} {109}},\ \bibinfo {pages} {195703} (\bibinfo
  {year} {2012})}\BibitemShut {NoStop}%
\bibitem [{\citenamefont {Appert-Rolland}\ \emph {et~al.}(2008)\citenamefont
  {Appert-Rolland}, \citenamefont {Derrida}, \citenamefont {Lecomte},\ and\
  \citenamefont {van Wijland}}]{Appert2008}%
  \BibitemOpen
  \bibfield  {author} {\bibinfo {author} {\bibfnamefont {C.}~\bibnamefont
  {Appert-Rolland}}, \bibinfo {author} {\bibfnamefont {B.}~\bibnamefont
  {Derrida}}, \bibinfo {author} {\bibfnamefont {V.}~\bibnamefont {Lecomte}}, \
  and\ \bibinfo {author} {\bibfnamefont {F.}~\bibnamefont {van Wijland}},\
  }\href@noop {} {\bibfield  {journal} {\bibinfo  {journal} {Phys. Rev. E}\
  }\textbf {\bibinfo {volume} {78}},\ \bibinfo {pages} {021122} (\bibinfo
  {year} {2008})}\BibitemShut {NoStop}%
\bibitem [{\citenamefont {Hurtado}\ and\ \citenamefont
  {Garrido}(2011)}]{Hurtado2011}%
  \BibitemOpen
  \bibfield  {author} {\bibinfo {author} {\bibfnamefont {P.~I.}\ \bibnamefont
  {Hurtado}}\ and\ \bibinfo {author} {\bibfnamefont {P.~L.}\ \bibnamefont
  {Garrido}},\ }\href@noop {} {\bibfield  {journal} {\bibinfo  {journal} {Phys.
  Rev. Lett.}\ }\textbf {\bibinfo {volume} {107}},\ \bibinfo {pages} {180601}
  (\bibinfo {year} {2011})}\BibitemShut {NoStop}%
\bibitem [{\citenamefont {Espigares}\ \emph {et~al.}(2013)\citenamefont
  {Espigares}, \citenamefont {Garrido},\ and\ \citenamefont
  {Hurtado}}]{Espigares2013}%
  \BibitemOpen
  \bibfield  {author} {\bibinfo {author} {\bibfnamefont {C.~P.}\ \bibnamefont
  {Espigares}}, \bibinfo {author} {\bibfnamefont {P.~L.}\ \bibnamefont
  {Garrido}}, \ and\ \bibinfo {author} {\bibfnamefont {P.~I.}\ \bibnamefont
  {Hurtado}},\ }\href@noop {} {\bibfield  {journal} {\bibinfo  {journal} {Phys.
  Rev. E}\ }\textbf {\bibinfo {volume} {87}},\ \bibinfo {pages} {032115}
  (\bibinfo {year} {2013})}\BibitemShut {NoStop}%
\bibitem [{\citenamefont {Jack}\ \emph {et~al.}(2015)\citenamefont {Jack},
  \citenamefont {Thompson},\ and\ \citenamefont {Sollich}}]{Jack2015}%
  \BibitemOpen
  \bibfield  {author} {\bibinfo {author} {\bibfnamefont {R.~L.}\ \bibnamefont
  {Jack}}, \bibinfo {author} {\bibfnamefont {I.~R.}\ \bibnamefont {Thompson}},
  \ and\ \bibinfo {author} {\bibfnamefont {P.}~\bibnamefont {Sollich}},\
  }\href@noop {} {\bibfield  {journal} {\bibinfo  {journal} {Phys. Rev. Lett.}\
  }\textbf {\bibinfo {volume} {114}},\ \bibinfo {pages} {060601} (\bibinfo
  {year} {2015})}\BibitemShut {NoStop}%
\bibitem [{\citenamefont {Karevski}\ and\ \citenamefont
  {Sch\"utz}(2017)}]{Karevski2017}%
  \BibitemOpen
  \bibfield  {author} {\bibinfo {author} {\bibfnamefont {D.}~\bibnamefont
  {Karevski}}\ and\ \bibinfo {author} {\bibfnamefont {G.~M.}\ \bibnamefont
  {Sch\"utz}},\ }\href@noop {} {\bibfield  {journal} {\bibinfo  {journal}
  {Phys. Rev. Lett.}\ }\textbf {\bibinfo {volume} {118}},\ \bibinfo {pages}
  {030601} (\bibinfo {year} {2017})}\BibitemShut {NoStop}%
\bibitem [{\citenamefont {Baek}\ \emph {et~al.}(2017)\citenamefont {Baek},
  \citenamefont {Kafri},\ and\ \citenamefont {Lecomte}}]{Baek2017}%
  \BibitemOpen
  \bibfield  {author} {\bibinfo {author} {\bibfnamefont {Y.}~\bibnamefont
  {Baek}}, \bibinfo {author} {\bibfnamefont {Y.}~\bibnamefont {Kafri}}, \ and\
  \bibinfo {author} {\bibfnamefont {V.}~\bibnamefont {Lecomte}},\ }\href@noop
  {} {\bibfield  {journal} {\bibinfo  {journal} {Phys. Rev. Lett.}\ }\textbf
  {\bibinfo {volume} {118}},\ \bibinfo {pages} {030604} (\bibinfo {year}
  {2017})}\BibitemShut {NoStop}%
\bibitem [{\citenamefont {Vaikuntanathan}\ \emph {et~al.}(2014)\citenamefont
  {Vaikuntanathan}, \citenamefont {Gingrich},\ and\ \citenamefont
  {Geissler}}]{Vaikuntanathan2014}%
  \BibitemOpen
  \bibfield  {author} {\bibinfo {author} {\bibfnamefont {S.}~\bibnamefont
  {Vaikuntanathan}}, \bibinfo {author} {\bibfnamefont {T.~R.}\ \bibnamefont
  {Gingrich}}, \ and\ \bibinfo {author} {\bibfnamefont {P.~L.}\ \bibnamefont
  {Geissler}},\ }\href@noop {} {\bibfield  {journal} {\bibinfo  {journal}
  {Phys. Rev. E}\ }\textbf {\bibinfo {volume} {89}},\ \bibinfo {pages} {062108}
  (\bibinfo {year} {2014})}\BibitemShut {NoStop}%
\bibitem [{\citenamefont {Weber}\ \emph {et~al.}(2015)\citenamefont {Weber},
  \citenamefont {Shukla},\ and\ \citenamefont {Pande}}]{Weber2015}%
  \BibitemOpen
  \bibfield  {author} {\bibinfo {author} {\bibfnamefont {J.~K.}\ \bibnamefont
  {Weber}}, \bibinfo {author} {\bibfnamefont {D.}~\bibnamefont {Shukla}}, \
  and\ \bibinfo {author} {\bibfnamefont {V.~S.}\ \bibnamefont {Pande}},\
  }\href@noop {} {\bibfield  {journal} {\bibinfo  {journal} {Proc. Natl. Acad.
  Sci. USA}\ }\textbf {\bibinfo {volume} {112}},\ \bibinfo {pages} {10377}
  (\bibinfo {year} {2015})}\BibitemShut {NoStop}%
\bibitem [{\citenamefont {Weber}\ \emph {et~al.}(2013)\citenamefont {Weber},
  \citenamefont {Jack},\ and\ \citenamefont {Pande}}]{Weber2013}%
  \BibitemOpen
  \bibfield  {author} {\bibinfo {author} {\bibfnamefont {J.~K.}\ \bibnamefont
  {Weber}}, \bibinfo {author} {\bibfnamefont {R.~L.}\ \bibnamefont {Jack}}, \
  and\ \bibinfo {author} {\bibfnamefont {V.~S.}\ \bibnamefont {Pande}},\
  }\href@noop {} {\bibfield  {journal} {\bibinfo  {journal} {J. Am. Chem.
  Soc.}\ }\textbf {\bibinfo {volume} {135}},\ \bibinfo {pages} {5501} (\bibinfo
  {year} {2013})}\BibitemShut {NoStop}%
\bibitem [{\citenamefont {Mey}\ \emph {et~al.}(2014)\citenamefont {Mey},
  \citenamefont {Geissler},\ and\ \citenamefont {Garrahan}}]{Mey2014}%
  \BibitemOpen
  \bibfield  {author} {\bibinfo {author} {\bibfnamefont {A.~S. J.~S.}\
  \bibnamefont {Mey}}, \bibinfo {author} {\bibfnamefont {P.~L.}\ \bibnamefont
  {Geissler}}, \ and\ \bibinfo {author} {\bibfnamefont {J.~P.}\ \bibnamefont
  {Garrahan}},\ }\href@noop {} {\bibfield  {journal} {\bibinfo  {journal}
  {Phys. Rev. E}\ }\textbf {\bibinfo {volume} {89}},\ \bibinfo {pages} {032109}
  (\bibinfo {year} {2014})}\BibitemShut {NoStop}%
\bibitem [{\citenamefont {Garrahan}(2018)}]{Garrahan2018}%
  \BibitemOpen
  \bibfield  {author} {\bibinfo {author} {\bibfnamefont {J.~P.}\ \bibnamefont
  {Garrahan}},\ }\href@noop {} {\bibfield  {journal} {\bibinfo  {journal}
  {Physica A}\ }\textbf {\bibinfo {volume} {504}},\ \bibinfo {pages} {130}
  (\bibinfo {year} {2018})}\BibitemShut {NoStop}%
\bibitem [{\citenamefont {Ates}\ \emph {et~al.}(2012)\citenamefont {Ates},
  \citenamefont {Olmos}, \citenamefont {Garrahan},\ and\ \citenamefont
  {Lesanovsky}}]{Ates2012}%
  \BibitemOpen
  \bibfield  {author} {\bibinfo {author} {\bibfnamefont {C.}~\bibnamefont
  {Ates}}, \bibinfo {author} {\bibfnamefont {B.}~\bibnamefont {Olmos}},
  \bibinfo {author} {\bibfnamefont {J.~P.}\ \bibnamefont {Garrahan}}, \ and\
  \bibinfo {author} {\bibfnamefont {I.}~\bibnamefont {Lesanovsky}},\
  }\href@noop {} {\bibfield  {journal} {\bibinfo  {journal} {Phys. Rev. A}\
  }\textbf {\bibinfo {volume} {85}},\ \bibinfo {pages} {043620} (\bibinfo
  {year} {2012})}\BibitemShut {NoStop}%
\bibitem [{\citenamefont {Lesanovsky}\ \emph {et~al.}(2013)\citenamefont
  {Lesanovsky}, \citenamefont {van Horssen}, \citenamefont {Guta},\ and\
  \citenamefont {Garrahan}}]{Lesanovsky2013}%
  \BibitemOpen
  \bibfield  {author} {\bibinfo {author} {\bibfnamefont {I.}~\bibnamefont
  {Lesanovsky}}, \bibinfo {author} {\bibfnamefont {M.}~\bibnamefont {van
  Horssen}}, \bibinfo {author} {\bibfnamefont {M.}~\bibnamefont {Guta}}, \ and\
  \bibinfo {author} {\bibfnamefont {J.~P.}\ \bibnamefont {Garrahan}},\
  }\href@noop {} {\bibfield  {journal} {\bibinfo  {journal} {Phys. Rev. Lett.}\
  }\textbf {\bibinfo {volume} {110}},\ \bibinfo {pages} {150401} (\bibinfo
  {year} {2013})}\BibitemShut {NoStop}%
\bibitem [{Note1()}]{Note1}%
  \BibitemOpen
  \bibinfo {note} {We indicate operators in the system with capital letters
  such as $J$, and super-operators on the system by calligraphic font symbols
  such as $\protect \mathcal {V}$. We use bold symbols such as ${\protect
  \mathbf V}$ and blackboard ones such as ${\protect \mathbb T}$ to indicate
  operators and super-operators, respectively, in the system-controller
  combined system of Sec.\ IV.B.}\BibitemShut {Stop}%
\bibitem [{\citenamefont {Lindblad}(1976)}]{Lindblad1976}%
  \BibitemOpen
  \bibfield  {author} {\bibinfo {author} {\bibfnamefont {G.}~\bibnamefont
  {Lindblad}},\ }\href@noop {} {\bibfield  {journal} {\bibinfo  {journal}
  {Comm. Math. Phys}\ }\textbf {\bibinfo {volume} {48}},\ \bibinfo {pages}
  {119} (\bibinfo {year} {1976})}\BibitemShut {NoStop}%
\bibitem [{\citenamefont {Gorini}\ \emph {et~al.}(1976)\citenamefont {Gorini},
  \citenamefont {Kossakowski},\ and\ \citenamefont {Sudarshan}}]{Gorini1976}%
  \BibitemOpen
  \bibfield  {author} {\bibinfo {author} {\bibfnamefont {V.}~\bibnamefont
  {Gorini}}, \bibinfo {author} {\bibfnamefont {A.}~\bibnamefont {Kossakowski}},
  \ and\ \bibinfo {author} {\bibfnamefont {E.~C.~G.}\ \bibnamefont
  {Sudarshan}},\ }\href@noop {} {\bibfield  {journal} {\bibinfo  {journal} {J.
  Mat. Phys.}\ }\textbf {\bibinfo {volume} {17}},\ \bibinfo {pages} {821}
  (\bibinfo {year} {1976})}\BibitemShut {NoStop}%
\bibitem [{\citenamefont {Nagourney}\ \emph {et~al.}(1986)\citenamefont
  {Nagourney}, \citenamefont {Sandberg},\ and\ \citenamefont
  {Dehmelt}}]{Nagourney1986}%
  \BibitemOpen
  \bibfield  {author} {\bibinfo {author} {\bibfnamefont {W.}~\bibnamefont
  {Nagourney}}, \bibinfo {author} {\bibfnamefont {J.}~\bibnamefont {Sandberg}},
  \ and\ \bibinfo {author} {\bibfnamefont {H.}~\bibnamefont {Dehmelt}},\
  }\href@noop {} {\bibfield  {journal} {\bibinfo  {journal} {Phys. Rev. Lett.}\
  }\textbf {\bibinfo {volume} {56}},\ \bibinfo {pages} {2797} (\bibinfo {year}
  {1986})}\BibitemShut {NoStop}%
\bibitem [{\citenamefont {Sauter}\ \emph {et~al.}(1986)\citenamefont {Sauter},
  \citenamefont {Neuhauser}, \citenamefont {Blatt},\ and\ \citenamefont
  {Toschek}}]{Sauter1986}%
  \BibitemOpen
  \bibfield  {author} {\bibinfo {author} {\bibfnamefont {T.}~\bibnamefont
  {Sauter}}, \bibinfo {author} {\bibfnamefont {W.}~\bibnamefont {Neuhauser}},
  \bibinfo {author} {\bibfnamefont {R.}~\bibnamefont {Blatt}}, \ and\ \bibinfo
  {author} {\bibfnamefont {P.}~\bibnamefont {Toschek}},\ }\href@noop {}
  {\bibfield  {journal} {\bibinfo  {journal} {Phys. Rev. Lett.}\ }\textbf
  {\bibinfo {volume} {57}},\ \bibinfo {pages} {1696} (\bibinfo {year}
  {1986})}\BibitemShut {NoStop}%
\bibitem [{\citenamefont {Bergquist}\ \emph {et~al.}(1986)\citenamefont
  {Bergquist}, \citenamefont {Hulet}, \citenamefont {Itano},\ and\
  \citenamefont {Wineland}}]{Bergquist1986}%
  \BibitemOpen
  \bibfield  {author} {\bibinfo {author} {\bibfnamefont {J.}~\bibnamefont
  {Bergquist}}, \bibinfo {author} {\bibfnamefont {R.~G.}\ \bibnamefont
  {Hulet}}, \bibinfo {author} {\bibfnamefont {W.~M.}\ \bibnamefont {Itano}}, \
  and\ \bibinfo {author} {\bibfnamefont {D.}~\bibnamefont {Wineland}},\
  }\href@noop {} {\bibfield  {journal} {\bibinfo  {journal} {Phys. Rev. Lett.}\
  }\textbf {\bibinfo {volume} {57}},\ \bibinfo {pages} {1699} (\bibinfo {year}
  {1986})}\BibitemShut {NoStop}%
\bibitem [{\citenamefont {Budini}\ \emph {et~al.}(2014)\citenamefont {Budini},
  \citenamefont {Turner},\ and\ \citenamefont {Garrahan}}]{Budini2014}%
  \BibitemOpen
  \bibfield  {author} {\bibinfo {author} {\bibfnamefont {A.~A.}\ \bibnamefont
  {Budini}}, \bibinfo {author} {\bibfnamefont {R.~M.}\ \bibnamefont {Turner}},
  \ and\ \bibinfo {author} {\bibfnamefont {J.~P.}\ \bibnamefont {Garrahan}},\
  }\href@noop {} {\bibfield  {journal} {\bibinfo  {journal} {J. Stat. Mech.}\
  ,\ \bibinfo {pages} {P03012}} (\bibinfo {year} {2014})}\BibitemShut {NoStop}%
\bibitem [{\citenamefont {Kiukas}\ \emph {et~al.}(2015)\citenamefont {Kiukas},
  \citenamefont {Guta}, \citenamefont {Lesanovsky},\ and\ \citenamefont
  {Garrahan}}]{Kiukas2015}%
  \BibitemOpen
  \bibfield  {author} {\bibinfo {author} {\bibfnamefont {J.}~\bibnamefont
  {Kiukas}}, \bibinfo {author} {\bibfnamefont {M.}~\bibnamefont {Guta}},
  \bibinfo {author} {\bibfnamefont {I.}~\bibnamefont {Lesanovsky}}, \ and\
  \bibinfo {author} {\bibfnamefont {J.~P.}\ \bibnamefont {Garrahan}},\
  }\href@noop {} {\bibfield  {journal} {\bibinfo  {journal} {Phys. Rev. E}\
  }\textbf {\bibinfo {volume} {92}},\ \bibinfo {pages} {012132} (\bibinfo
  {year} {2015})}\BibitemShut {NoStop}%
\bibitem [{\citenamefont {Breuer}(2004)}]{Breuer2004}%
  \BibitemOpen
  \bibfield  {author} {\bibinfo {author} {\bibfnamefont {H.-P.}\ \bibnamefont
  {Breuer}},\ }\href@noop {} {\bibfield  {journal} {\bibinfo  {journal} {Phys.
  Rev. A}\ }\textbf {\bibinfo {volume} {70}},\ \bibinfo {pages} {012106}
  (\bibinfo {year} {2004})}\BibitemShut {NoStop}%
\bibitem [{\citenamefont {Hush}\ \emph {et~al.}(2015)\citenamefont {Hush},
  \citenamefont {Lesanovsky},\ and\ \citenamefont {Garrahan}}]{Hush2015}%
  \BibitemOpen
  \bibfield  {author} {\bibinfo {author} {\bibfnamefont {M.~R.}\ \bibnamefont
  {Hush}}, \bibinfo {author} {\bibfnamefont {I.}~\bibnamefont {Lesanovsky}}, \
  and\ \bibinfo {author} {\bibfnamefont {J.~P.}\ \bibnamefont {Garrahan}},\
  }\href@noop {} {\bibfield  {journal} {\bibinfo  {journal} {Phys. Rev. A}\
  }\textbf {\bibinfo {volume} {91}},\ \bibinfo {pages} {032113} (\bibinfo
  {year} {2015})}\BibitemShut {NoStop}%
\bibitem [{\citenamefont {Mabuchi}(2011)}]{Mabuchi2011}%
  \BibitemOpen
  \bibfield  {author} {\bibinfo {author} {\bibfnamefont {H.}~\bibnamefont
  {Mabuchi}},\ }\href@noop {} {\bibfield  {journal} {\bibinfo  {journal} {App.
  Phys. Lett.}\ }\textbf {\bibinfo {volume} {98}},\ \bibinfo {pages} {193109}
  (\bibinfo {year} {2011})}\BibitemShut {NoStop}%
\bibitem [{\citenamefont {Wu}\ and\ \citenamefont {Mabuchi}(2018)}]{Wu2018}%
  \BibitemOpen
  \bibfield  {author} {\bibinfo {author} {\bibfnamefont {J.}~\bibnamefont
  {Wu}}\ and\ \bibinfo {author} {\bibfnamefont {H.}~\bibnamefont {Mabuchi}},\
  }\href@noop {} {\bibfield  {journal} {\bibinfo  {journal} {Phys. Rev. A}\
  }\textbf {\bibinfo {volume} {98}},\ \bibinfo {pages} {013812} (\bibinfo
  {year} {2018})}\BibitemShut {NoStop}%
\bibitem [{\citenamefont {Chetrite}\ and\ \citenamefont
  {Touchette}(2013)}]{Chetrite2013}%
  \BibitemOpen
  \bibfield  {author} {\bibinfo {author} {\bibfnamefont {R.}~\bibnamefont
  {Chetrite}}\ and\ \bibinfo {author} {\bibfnamefont {H.}~\bibnamefont
  {Touchette}},\ }\href@noop {} {\bibfield  {journal} {\bibinfo  {journal}
  {Phys. Rev. Lett.}\ }\textbf {\bibinfo {volume} {111}},\ \bibinfo {pages}
  {120601} (\bibinfo {year} {2013})}\BibitemShut {NoStop}%
\bibitem [{\citenamefont {Lind}\ and\ \citenamefont
  {Marcus}(1995)}]{LindMarcus}%
  \BibitemOpen
  \bibfield  {author} {\bibinfo {author} {\bibfnamefont {D.}~\bibnamefont
  {Lind}}\ and\ \bibinfo {author} {\bibfnamefont {B.}~\bibnamefont {Marcus}},\
  }\href@noop {} {\emph {\bibinfo {title} {Symbolic Dynamics and Coding}}}\
  (\bibinfo  {publisher} {Cambridge University Press},\ \bibinfo {year}
  {1995})\BibitemShut {NoStop}%
\bibitem [{\citenamefont {Gough}\ and\ \citenamefont
  {James}(2009)}]{Gough2009}%
  \BibitemOpen
  \bibfield  {author} {\bibinfo {author} {\bibfnamefont {J.}~\bibnamefont
  {Gough}}\ and\ \bibinfo {author} {\bibfnamefont {M.~R.}\ \bibnamefont
  {James}},\ }\href@noop {} {\bibfield  {journal} {\bibinfo  {journal} {IEEE
  Transactions on Automatic Control}\ }\textbf {\bibinfo {volume} {54}},\
  \bibinfo {pages} {2530} (\bibinfo {year} {2009})}\BibitemShut {NoStop}%
\bibitem [{\citenamefont {Jack}\ and\ \citenamefont
  {Sollich}(2010)}]{Jack2010}%
  \BibitemOpen
  \bibfield  {author} {\bibinfo {author} {\bibfnamefont {R.~L.}\ \bibnamefont
  {Jack}}\ and\ \bibinfo {author} {\bibfnamefont {P.}~\bibnamefont {Sollich}},\
  }\href@noop {} {\bibfield  {journal} {\bibinfo  {journal} {Prog. Th. Phys.
  Supp.}\ }\textbf {\bibinfo {volume} {184}},\ \bibinfo {pages} {304} (\bibinfo
  {year} {2010})}\BibitemShut {NoStop}%
\bibitem [{\citenamefont {Chetrite}\ and\ \citenamefont
  {Touchette}(2015{\natexlab{a}})}]{Chetrite2015}%
  \BibitemOpen
  \bibfield  {author} {\bibinfo {author} {\bibfnamefont {R.}~\bibnamefont
  {Chetrite}}\ and\ \bibinfo {author} {\bibfnamefont {H.}~\bibnamefont
  {Touchette}},\ }\href@noop {} {\bibfield  {journal} {\bibinfo  {journal}
  {Ann. Henri Poincar\'e}\ }\textbf {\bibinfo {volume} {16}},\ \bibinfo {pages}
  {2005} (\bibinfo {year} {2015}{\natexlab{a}})}\BibitemShut {NoStop}%
\bibitem [{\citenamefont {Garrahan}(2016)}]{Garrahan2016}%
  \BibitemOpen
  \bibfield  {author} {\bibinfo {author} {\bibfnamefont {J.~P.}\ \bibnamefont
  {Garrahan}},\ }\href@noop {} {\bibfield  {journal} {\bibinfo  {journal} {J.
  Stat. Mech.}\ ,\ \bibinfo {pages} {073208}} (\bibinfo {year}
  {2016})}\BibitemShut {NoStop}%
\bibitem [{\citenamefont {Carollo}\ \emph {et~al.}(2018)\citenamefont
  {Carollo}, \citenamefont {Garrahan}, \citenamefont {Lesanovsky},\ and\
  \citenamefont {P{\'e}rez-Espigares}}]{Carollo2018}%
  \BibitemOpen
  \bibfield  {author} {\bibinfo {author} {\bibfnamefont {F.}~\bibnamefont
  {Carollo}}, \bibinfo {author} {\bibfnamefont {J.~P.}\ \bibnamefont
  {Garrahan}}, \bibinfo {author} {\bibfnamefont {I.}~\bibnamefont
  {Lesanovsky}}, \ and\ \bibinfo {author} {\bibfnamefont {C.}~\bibnamefont
  {P{\'e}rez-Espigares}},\ }\href@noop {} {\bibfield  {journal} {\bibinfo
  {journal} {Phys. Rev. A}\ }\textbf {\bibinfo {volume} {98}},\ \bibinfo
  {pages} {010103} (\bibinfo {year} {2018})}\BibitemShut {NoStop}%
\bibitem [{\citenamefont {Jack}\ and\ \citenamefont
  {Sollich}(2015)}]{Jack2015b}%
  \BibitemOpen
  \bibfield  {author} {\bibinfo {author} {\bibfnamefont {R.~L.}\ \bibnamefont
  {Jack}}\ and\ \bibinfo {author} {\bibfnamefont {P.}~\bibnamefont {Sollich}},\
  }\href@noop {} {\bibfield  {journal} {\bibinfo  {journal} {Euro. Phys. J.
  Spec. Topics}\ }\textbf {\bibinfo {volume} {224}},\ \bibinfo {pages} {2351}
  (\bibinfo {year} {2015})}\BibitemShut {NoStop}%
\bibitem [{\citenamefont {Chetrite}\ and\ \citenamefont
  {Touchette}(2015{\natexlab{b}})}]{Chetrite2015b}%
  \BibitemOpen
  \bibfield  {author} {\bibinfo {author} {\bibfnamefont {R.}~\bibnamefont
  {Chetrite}}\ and\ \bibinfo {author} {\bibfnamefont {H.}~\bibnamefont
  {Touchette}},\ }\href@noop {} {\bibfield  {journal} {\bibinfo  {journal} {J.
  Stat. Mech.}\ ,\ \bibinfo {pages} {P12001}} (\bibinfo {year}
  {2015}{\natexlab{b}})}\BibitemShut {NoStop}%
\bibitem [{\citenamefont {Evans}\ and\ \citenamefont
  {Majumdar}(2011)}]{evans2011}%
  \BibitemOpen
  \bibfield  {author} {\bibinfo {author} {\bibfnamefont {M.~R.}\ \bibnamefont
  {Evans}}\ and\ \bibinfo {author} {\bibfnamefont {S.~N.}\ \bibnamefont
  {Majumdar}},\ }\href {\doibase 10.1103/PhysRevLett.106.160601} {\bibfield
  {journal} {\bibinfo  {journal} {Phys. Rev. Lett.}\ }\textbf {\bibinfo
  {volume} {106}},\ \bibinfo {pages} {160601} (\bibinfo {year}
  {2011})}\BibitemShut {NoStop}%
\bibitem [{\citenamefont {Evans}\ and\ \citenamefont
  {Majumdar}(2014)}]{evans2014}%
  \BibitemOpen
  \bibfield  {author} {\bibinfo {author} {\bibfnamefont {M.~R.}\ \bibnamefont
  {Evans}}\ and\ \bibinfo {author} {\bibfnamefont {S.~N.}\ \bibnamefont
  {Majumdar}},\ }\href {http://stacks.iop.org/1751-8121/47/i=28/a=285001}
  {\bibfield  {journal} {\bibinfo  {journal} {J. Phys. A: Math. Theor.}\
  }\textbf {\bibinfo {volume} {47}},\ \bibinfo {pages} {285001} (\bibinfo
  {year} {2014})}\BibitemShut {NoStop}%
\bibitem [{\citenamefont {Reuveni}(2016)}]{reuveni2016}%
  \BibitemOpen
  \bibfield  {author} {\bibinfo {author} {\bibfnamefont {S.}~\bibnamefont
  {Reuveni}},\ }\href {\doibase 10.1103/PhysRevLett.116.170601} {\bibfield
  {journal} {\bibinfo  {journal} {Phys. Rev. Lett.}\ }\textbf {\bibinfo
  {volume} {116}},\ \bibinfo {pages} {170601} (\bibinfo {year}
  {2016})}\BibitemShut {NoStop}%
\bibitem [{\citenamefont {Pal}\ and\ \citenamefont {Reuveni}(2017)}]{pal2017}%
  \BibitemOpen
  \bibfield  {author} {\bibinfo {author} {\bibfnamefont {A.}~\bibnamefont
  {Pal}}\ and\ \bibinfo {author} {\bibfnamefont {S.}~\bibnamefont {Reuveni}},\
  }\href {\doibase 10.1103/PhysRevLett.118.030603} {\bibfield  {journal}
  {\bibinfo  {journal} {Phys. Rev. Lett.}\ }\textbf {\bibinfo {volume} {118}},\
  \bibinfo {pages} {030603} (\bibinfo {year} {2017})}\BibitemShut {NoStop}%
\end{thebibliography}%

\end{document}